\documentclass{article}
\usepackage{graphicx}
\usepackage{amsmath}
\usepackage{amsthm}
\usepackage{amssymb}
\usepackage{url}
\usepackage{mathrsfs}
\usepackage{appendix}

\usepackage{latexsym}
\usepackage[noadjust]{cite}
\newcommand{\bma}{\left[\begin{matrix}}
\newcommand{\ema}{\end{matrix}\right]}
\newcommand{\be}{\begin{equation}}
\newcommand{\ee}{\end{equation}}
\newcommand{\BC}{\mathbb{C}}
\newcommand{\BR}{\mathbb{R}}
\newcommand{\BH}{\mathbb{H}}
\newcommand{\mk}{\mathcal{K}}
\newcommand{\bx}{\mathbf{x}}
\newcommand{\br}{\mathbf{r}}
\newcommand{\bX}{\mathbf{X}}
\newcommand{\ct}{\mathcal{\tau}}

\newcommand{\cN}{\mathcal{N}}
\newcommand{\cM}{\mathcal{M}}
\newcommand{\cZ}{\mathcal{Z}}
\newcommand{\cJ}{\mathcal{J}}

\newcommand{\cW}{\mathcal{W}}
\newcommand{\bR}{{\mathbb R}}
\newcommand{\tm}{{\cal T}}

\newcommand{\hp}{{\hat \phi}}

\newcommand{\diag}{\mathrm{diag}}
\newcommand{\Ee}[3]{#1^{#2}_{\phantom{#2}#3}}
\newcommand{\Eee}[4]{#1^{#2}_{\phantom{#2}#3#4}}
\font\ddpp=msbm10 scaled \magstep 1

\newtheorem{lemma}{Lemma}

\newtheorem{proposition}{Proposition}
\newtheorem{remark}{Remark}
\newtheorem{thm}{Theorem}
\begin{document}
\title{On Conformal Infinity and Compactifications of the Minkowski Space}
\author{Arkadiusz Jadczyk\footnote{E-mail address: arkadiusz.jadczyk@cict.fr}\smallskip \\ \emph{Center CAIROS}, \emph{Institut de Math\'{e}matiques de
    Toulouse}\\ \emph{Universit\'{e} Paul Sabatier, 31062 TOULOUSE CEDEX  9, France }}
\maketitle
\begin{abstract}
Using the standard Cayley transform and elementary tools it is  reiterated  that the conformal compactification of the Minkowski space involves not only the ``cone at infinity'' but also the 2-sphere that is at the base of this cone. We  represent this 2-sphere by two additionally marked points on the Penrose diagram for the compactified Minkowski space. Lacks and omissions in the existing literature are described, Penrose diagrams are derived  for both, simple compactification and  its double covering space, which is discussed in some detail using both the $U(2)$ approach and the exterior and Clifford algebra methods. Using the Hodge $\star$ operator twistors (i.e. vectors of the pseudo-Hermitian space $H_{2,2}$) are realized as spinors (i.e., vectors of a faithful irreducible representation of the even Clifford algebra) for the conformal group $SO(4,2)/Z_2.$ Killing vector fields corresponding to the left action of $U(2)$ on itself are explicitly calculated. Isotropic cones and corresponding projective quadrics in  $H_{p,q}$ are also discussed. Applications to flat conformal structures, including the normal Cartan connection and conformal development has been discussed in some detail.
\end{abstract}
\section{Introduction}
The term {\em compactification\,} can have several different meanings. Given a manifold $\cM$ we may try to embed it into a compact one and take its closure. Or, we can attach to $\cM$ ideal boundary points or boundary components so as to obtain a compact space. In physics compactification of space--time can be used either in order to study its conformal invariance, or to study its asymptotic flatness, or its singularities. In the available literature the differences between these different approaches are not always made clear and the mathematical language involved is not always as precise as one would wish.

This paper is a compromise between being completely self--contained and a typical specialized article. We use techniques of algebra and geometry but we avoid twistor notation of Penrose school which can be confusing to many mathematicians. The paper is aimed at mathematicians interested in mathematical properties of Minkowski space related to projective geometry, and at mathematical physicists interested in the subject. Relativists will find next to nothing of interest for them in the material below (perhaps except of a warning about how errors can easily propagate). They have their own aims and techniques and, as a rule, are usually not interested in generalizations going beyond four space--time dimensions.

In section \ref{sec:ccms} we review the conformal compactification ${\tilde M}=U(2)$ of the Minkowski space $M.$ We are following there the elegant and simple method of A. Uhlmann \cite{uhlmann63} by using $2\times 2$ matrices and the Cayley transform. We are also investigating in some detail the structure of the ``light cone at infinity'', that is the set difference ${\tilde M}\setminus M$ and point out that it consists not only of the (double) light cone, but also of a 2-sphere that connects the two cones - a fact that was known to Roger Penrose \cite[p. 178]{penrin}. This fact was not always realized by other authors writing on this subject even when they quoted Penrose (cf. e.g., Sec. \ref{sec:tod}). Additionally, as a complement to this particular representation of ${\tilde M},$ in appendix \ref{app:app1}, we calculate vector fields on $M$ corresponding to one--parameter subgroups of $U(2)$ acting on itself by left translations.

In section \ref{sec:tod}, as an educational example, we discuss in some detail the faulty argument and the missing 2-sphere in   \cite{tod}. In particular we reproduce a crucial part of reasoning used in \cite{tod} and point out the omission explicitly. Similar omissions, this time taken from \cite{akivis} and also from a recent papers on conformal field theory, are discussed in section \ref{sec:akivis}.

In section \ref{sec:pn}, geometrical representation of the conformal compactification ${\tilde M}$ is discussed using the cylinder representation of Einstein's static universe - the standard representation in general relativity. This leads to a two--dimensional diagram - a version of the Penrose diagram (cf. Fig. \ref{fig:fig1}), with the two 2-spheres that need to be identified. Owing to this identification no intrinsic distinction between $\cJ^+$ and $\cJ^-$ is possible. In Fig. \ref{fig:fig3} and Fig. \ref{fig:fig4} we mark these two parts of the conformal infinity in order to be able to compare this diagram with those (as in Fig. \ref{fig:fig3a}) found in the standard literature.

In section \ref{sec:il}, the explicit action of the Poincar\'{e} group on the conformal infinity is calculated, where it is in particular shown that this action is transitive there. A lack of a mathematical precision in the mathematical literature on the subject is also elucidated.

Section \ref{sec:geo} starts with a simple exercise showing a geometrically amusing fact that null geodesics can be completely trapped at infinity. A role of the conformal inversion, and the signature of the induced metric is also discussed there. Then, a pictorial representation of the infinity is given, first as a double cone with identified vertices in Fig. \ref{fig:fig3}, then, more correct as far as its differentiability properties are concerned,  as a squeezed torus in Fig. \ref{fig:fig4}. A typical, almost identically looking, but with a different meaning, picture - taken from   \cite{beem} - is shown in Fig. \ref{fig:fig3a}. The squeeze point in Fig. \ref{fig:fig4} corresponds to what is usually denoted as $I^0,I^+,I^-$ (or $i^0,i^+,i^-$) in the standard literature. All three points coincide in our case.\footnote{A. Uhlmann \cite{uhlmann63} conjectured that it may be a squeezed Klein's bottle. Klein's bottle is unnecessary as long as we do not care about the embedding. Squeezed torus does the job.} A correct image, which we reproduce here in Fig. \ref{fig:fig3b} can be found in Fig. 2 of  \cite{flores}. It may be worth quoting the following remarks from the monograph of Penrose and Rindler \cite[p. 298]{penrin}:
\begin{quotation}
``Having this natural association between the points of $\cJ^-$ and $\cJ^+$, for Minkowski space, it is in some respect natural to make identification between $\cJ^-$ and $\cJ^+$, the point $A^-$ being identified with $A^+$ and $\cJ^-$ and $\cJ^+$ written as $\cJ.$ If we do this, then, for the sake of continuity we should also identify $I^-$ with $I^0,$ and $I^0$ with $I^+.$''
\end{quotation}
To which they added:
\begin{quotation}
``For reasons that we shall see in more detail later, such identification cannot be satisfactorily carried out in {\em curved\,} asymptotically flat spaces. (Not only is there apparently no {\em canonical\,} way of performing such identifications in general, but, when the total mass is non--zero {\em any\,} identification would lead to failure of the required regularity conditions along the identification hypersurface.) For many purposes, the identification of $\cJ^-$ with $\cJ^+$ may, even in Minkowski space, seem unphysical (and, of course, it need not be made). However, for various mathematical purposes the identification is very useful...''
\end{quotation}
In subsection \ref{sec:dc} we discuss the double cover of ${\tilde M},$ that can be obtained by the same method as in section \ref{sec:tod} but by considering positive rays rather than generator lines.\footnote{This construction is also briefly mentioned in   \cite[p. 180]{lernercmp}. It is also worthwhile to mention that $(U(1)\times SU(2))/Z_2,\, Z_2=\{I,-I\},$ with the topology of $(S^1\times S^3)/Z_2$ is homeomorphic, as a manifold, to its double cover $U(1)\times SU(2)$ - cf.  \cite{chevlie} and \cite{levichev}.}. This leads us to the compactification with the past infinity $\cJ^-$ and future infinity $\cJ^+$ different, but $I^-$ and $I^+$ are identified, though different from $I^0.$ The resulting Penrose diagram is given in Fig. \ref{fig:fig2}, and the ensuing graphic representation of the conformal infinity is pictured in Fig. \ref{fig:fig5} and in Fig. \ref{fig:fig6}.\\
\noindent We follow here method used by Kopczy\'{n}ski and Woronowicz in   \cite{wk}, but this time applied to the double cover of $M.$ Moreover, we identify the antilinear map $x\mapsto x^\perp$  used by these authors as a Hodge $\star$ operator adapted for a complex vector space $V$ equipped with a non--degenerate sesquilinear form\footnote{For a discussion  in case of positive definite scalar product cf. e.g., \cite{sto}.}. After a general introduction, for an arbitrary signature, starting with the Grassmann algebra endowed with the natural scalar product, we specialize to the case of  signature $(2,2),$ $V\approx H_{2,2},$ and relate the two compactification methods - one in which the points of the double covering of the compactified Minkowski space are represented by oriented maximal isotropic subspaces of a four dimensional complex space endowed with a sesquilinear form of signature $(2,2),$ and the one discussed in Sec. \ref{sec:dc} based on rays of the null cone in $6$-dimensional real space endowed with a scalar product with signature $(4,2).$ We derive explicit formulas connecting the $U(2)$ compactification and the one based on $H_{2,2}.$

In order to show how the compactified Minkowski space enters more general conformal structures on manifolds, in section \ref{sec:fcs} we briefly review geometry of conformal structures, second-order frames and the normal Cartan connection. We end this section by explicitly calculating the standard embedding of Minkowski space into the compact projective hyperquadric using the conformal development.
\section{Conformally compactified Minkowski space\label{sec:ccms}}
In this section we follow idea of Armin Uhlmann \cite{uhlmann63}.
Let $H(2)$ be the real vector space of complex $2\times 2$ Hermitian matrices. Let $M$ be the Minkowski space endowed with the standard coordinates $x^1,x^2,x^3,x^0,$\footnote{Sometimes, as an alternative, will set $x^0=x^4,$ and write $x=(x^1,...,x^4)\in M.$} and the quadratic form $q(x)=-(x^0)^2+(x^1)^2+(x^2)^2+(x^3)^2,$ and let $\varphi: M\rightarrow H(2)$ be the isomorphism given by\footnote{Cf. e.g.,   \cite[p. 324]{deheuvels}.}
\be \varphi (x) = X = \bma x^0+x^3&x^1-ix^2\\x^1+ix^2&x^0-x^3\ema.\label{eq:U}\ee
Then we have
\be \det (X)=(x^0)^2-\left((x^1)^2+(x^2)^2+(x^3)^2\right)=-q(x).\label{eq:det}\ee

Let $U(2)$ be the group of all unitary $2\times 2$ matrices with complex entries.
Let $u: H(2)\rightarrow U(2)$ be the Cayley transform:
$$ u(X)=U=\frac{X-iI}{X+iI}.$$

Notice that, because of $X$ being Hermitian,   $\det (X+iI)\neq 0.$ We then have
\begin{eqnarray}
I+U &=&\frac{iI+X+X-iI}{X+iI}=\frac{2X}{X+iI},\nonumber\\
I-U&=&\frac{X+iI-X+iI}{X+iI}=\frac{2i}{X+iI}.
\label{eq:upo}\end{eqnarray}
In particular $\det (I-U)\neq 0$ and
\be
X=i\,\frac{I+U}{I-U}.\label{eq:ict}\ee
It easily follows that $\psi= u\circ \varphi : M\rightarrow U(2)$ is a bijection from $M$ onto the open subset of $U(2)$ consisting of those $U$ for which $\det (I-U)\neq 0.$
\begin{remark}It may be useful for the reader to see the explicit form of $\psi(x)$ for any $x\in M,$ namely
\be U=\psi(x)=\frac{1}{-q(x)-1+2ix^0}\bma 1-q(x)+2ix^3&2(ix^1+x^2)\\2(ix^1-x^2)&1-q(x)-2ix^3\ema.\label{eq:uu}\ee
We also have, explicitly:
\be \det (I-U)=\frac{4}{1+q(x)-2ix^0},\qquad \det (I+U)=\frac{4q(x)}{1+q(x)-2ix^0}.\label{eq:dets}\ee
The first one of the last two equalities shows that for any $U\in\psi(M),$ $\det(I-U)\neq 0,$ while the second one states that $\det(I+U)=0$ if and only if $q(x)=0.$ Notice that the quantity $1+q(x)-2ix^0\neq 0$ for all $x\in M.$
\end{remark}
\noindent Let us now determine the structure of the remaining set $\mathfrak{I}:$
$$ \mathfrak{I}=U(2)\setminus\psi(M)=\{U\in U(2):\, \det (I-U)=0\}.$$
Let $m:U(2)\rightarrow U(2)$ be the diffeomorphism of $U(2)$ given by
$ m(U)=-U,$
i.e., the group translation by $-I.$ Let us investigate the structure of the set $m(\mathfrak{I})$ - the image of $\mathfrak{I}\subset U(2)$ under $m.$ We split this set into two disjoint non empty components $\mathfrak{I}_c$ and $\mathfrak{I}_s$ defined by
$$\mathfrak{I}_c =m(\mathfrak{I})\setminus \mathfrak{I},\quad \mbox{and}\quad
\mathfrak{I}_s =m(\mathfrak{I})\cap \mathfrak{I}.$$
\begin{remark}To see that both sets, $\mathfrak{I}_c$ and $\mathfrak{I}_s,$ are non empty, notice that $U_0=-I=m(I)$ is not in $\mathfrak{I},$ but is in $m(\mathfrak{I}).$ Therefore $U_0$ is in $\mathfrak{I}_c.$ On the other hand let $U_1=\left(\begin{smallmatrix}1&0\\0&-1\end{smallmatrix}\right).$ Then $U_1$ and $-U_1=m(U_1)$ are in $\mathfrak{I},$ thus $U_1$ is in $\mathfrak{I}_s.$
\end{remark}
\noindent The set $\mathfrak{I}_c$ is, by its definition in the range of Cayley transform, therefore we can apply $\psi^{-1}$ to $\mathfrak{I}_c.$

Denoting by $K$ the light cone through the origin: $K=\{x\in M:\, q(x)=0\},$ let us show that
\be \psi^{-1}(\mathfrak{I}_c)=K.\label{eq:cc}\ee
With $x\in M$ we have that $x\in\psi^{-1}(\mathfrak{I}_c)$ if and only if $\psi(x)\in\mathfrak{I}_c,$ that is if and only if $(U\in m(\mathfrak{I} ))$ and $(U\not\in\mathfrak{I}).$ That is $x\in\psi^{-1}(\mathfrak{I}_c)$ if and only if $\det(I+U)=0$ and $\det(I-U)\neq 0.$ It follows now from Eq. (\ref{eq:dets}) that $\det(I-U)$ is automatically non--zero, and that $\det(I+U)=0$ is equivalent to $q(x)=0,$ that is $x\in K.$

It remains to identify the set $\mathfrak{I}_s.$ Let $j:\, U(2)\rightarrow U(2)$ be the map $ j(U)=iU,$
i.e., the translation by $i.$ It follows from the very definition that $U\in\mathfrak{I}_s$ is equivalent to: $\det(I-U)=0$ and $\det(I+U)=0.$ It follows that $U\in \mathfrak{I}_s$ if and only if one eigenvalue of $U$ is equal $+1$ while the other eigenvalue is equal $-1.$ It follows that $j(U)=iU$ has eigenvalues $+i$ and $-i.$ Therefore $I-iU$ is invertible and $U=\varphi(X),$ with $X$ given by Eq. (\ref{eq:ict}) and $U$ replaced by $iU$. It follows that $j(U)$ is in the range of $\psi.$ Thus we conclude that $j(\mathfrak{I}_s)\subset \psi(M).$

\noindent Let us show that $\psi^{-1}(j(\mathfrak{I}_s ))$ is the 2-sphere:
$$ \psi^{-1}(j(\mathfrak{I}_s ))=\{x\in M:\, x^0=0,\, (x^1)^2+(x^2)^2+(x^3)^2=1\}.$$
With $U\in \mathfrak{I}_s$ let $x=\psi^{-1}(j(U)).$ Then
$
\psi(x)=X=i\,\frac{I+iU}{I-iU}.$
It follows that $X$ has eigenvalues $i\frac{1+i}{1-i}=-1$ and $i\frac{1-i}{1+i}=1,$ which is equivalent to $\det(X)=-1$ and $\rm{tr}(X)=0.$ Now, from Eq. (\ref{eq:U}) it follows that $\rm{tr}(X)=0$
is equivalent to $x^0=0,$ and then $\det(X)=-1$ is equivalent to $(x^1)^2+(x^2)^2+(x^3)^2=1,$ which concludes our proof.

It follows from the above that $U(2)\setminus \psi(M)$ consists of two pieces. The first piece is the set of all unitary matrices with precisely one eigenvalue equal to $-1,$ the other eigenvalue different from $+1.$ This piece has the structure of the {\em light cone at infinity\,}. The matrix $U=-I$ is the apex of this cone. The second piece consists of unitary matrices with one eigenvalue equal to $-1,$ the other eigenvalue being $+1.$ This piece is the {\em 2-sphere at infinity\,} that forms ``a base'' of the light cone at infinity.

\begin{remark}
A closely related derivation of this fact can be found in   \cite[Theorem 6]{daigneault}. This pedagogical paper is closely related in spirit and is a recommended reading for all those interested in the subject.
\end{remark}
\begin{remark}It is easy to calculate the result of the transformation $x\mapsto x'$ corresponding to the left translation $U\mapsto iU=j(U).$ The result of a simple calculation reads:
$$
x^{0'}=\frac{1+q(x)}{1-q(x)-2x^0}\quad \mbox{and}\quad
\bx'=\frac{2\bx}{1-q(x)-2x^0}.
$$
This particular transformation can be interpreted in terms of conformal transformations $T(a)x=x+a,\, K(a)=RT(a)R,\, D(\lambda)x=\lambda x,$ where $R$ is the inversion $R(x)=x/q(x).$ A simple calculation shows that $$x'=T(-a)D(2)K(a)x,$$ where $a^0=-1,\,{\bf a}=0.$ The transformation is singular on the light cone centered at $-a.$
\end{remark}
In appendix \ref{app:app1} we calculate the conformal vector fields on Minkowski space corresponding to left actions of one--parameter subgroups of $U(2).$

\section{The overlooked 2-sphere\label{sec:tod}}
In their {\em Introduction to Twistor Theory\,} \cite[Chapt. 5]{tod}, {\em Compactified Minkowski Space\,}, the authors obtain their ``cone at infinity'' using a different method and, as we will see, their incomplete reasoning leads to their neglecting of the 2-sphere at infinity. First, we will reproduce their reasoning, using their notation, with slight changes, simplifications, and with some elucidating comments. Then, we will present our corrected derivation and its result.
\subsection{Reasoning of Huggett and Tod\label{sec:ref2}}
Here we will present the essence of the reasoning in   \cite{tod}, though with some changes of the notation.  We denote by $M$ the standard Minkowski space, that is $E^{3,1}=\BR^3\oplus\BR^1,$ with coordinates $x=(\bx,t),$ endowed with the quadratic form $q(x)=\bx^2-t^2, $ where $\bx=(x^1,x^2,x^3),$ and $\bx^2$ is the standard Euclidean quadratic form of $\BR^3:$ $\bx^2=(x^1)^2+(x^2)^2+(x^3)^2.$
Let $E^{1,1}$ be $\BR^2$ endowed with the quadratic form $q_2$ defined by $q_2(x_5,x_6)=(x^5)^2-(x^6)^2,\, (x_5,x_6)\in \BR^2.$
We denote by $E^{4,2}$ the $6$--dimensional space $E^{3,1}\oplus E^{1,1} ,$ with coordinates $(Z^\alpha)=(x,x_5,x_6),$ and endowed with the quadratic form
$Q(x,x_5,x_6)=q(x)+q_2(x_5,x_6).$ In order to simplify a bit the notation, let us set, in this section,
$$ x^5=v,\quad x^6=w.$$

Let $\cN$ be the null cone of $E^{4,2}$ minus the origin: \be \cN=\{Z\in E^{4,2}:\,Z\neq 0 \mbox{ and } Q(Z)=0\},\label{eq:qzz}\ee and let $P\cN$ be the set of its generators, that is the set of straight lines through
the origin in the directions nullifying $Q(Z).$ In other words $P\cN=\cN/\sim,$ where, for $Z,Z'\in \cN,$ $Z\sim Z'$ if and only if there exists a nonzero $\mu \in \BR$ such that $Z'=\mu Z.$ We denote by $\pi$ the projection $\pi:\cN\rightarrow P\cN.$ Then $P\cN,$ with its projective topology, is a compact projective quadric. $P\cN$ is called the {\em compactified Minkowski space\,}.

Consider now the following smooth map between manifolds: $\ct: M\rightarrow E^{4,2}$ given by the formula:
\be \ct(\bx ,t )=(\bx ,t,\frac{1}{2}(1-q(x)),-\frac{1}{2}(1+q(x))).\label{eq:tau}\ee
The map $\ct$ is evidently injective. Let $\cZ$ be the hyperplane in $E^{4,2}:$
\be \cZ=\{Z\in E^{4,2}:\, v-w=1\}.\label{eq:z}\ee
\begin{lemma} The image $\ct(M)$ in $E^{4,2}$ coincides with the intersection $\cN\cap\cZ$ of $\cN$ with $\cZ.$
\end{lemma}
\begin{proof}It is clear that $\ct(x)\neq 0,$ and it also follows by an easy calculation that $Q(\ct(x))=0.$ Evidently, from Eq. (\ref{eq:tau}), $\ct(x)$ is also in $\cZ.$ Conversely,
let $Z=(x,v,w)$ be in $\cN\cap \cZ.$ From $Q(Z)=Q(x,v,w)=0$ we get $q(x)+v^2-w^2=0.$ But $v^2-w^2=(v-w)(v+w)$ so that from $v-w=1$ it follows that $q(x)+v+w=0.$ Together with $v-w=1$ it implies $q(x)+2v=1$ or $v=\frac{1}{2}(1-q(x))$ and $w=-\frac{1}{2}(1+q(x)).$ It follows that $Z=\ct(X).$\end{proof}

From now on we will follow the arguments in   \cite[p. 36]{tod} step by step, skipping what is not essential and adapting to our notation.
\begin{quotation} ``On any generator of $\cN$ with $v-w\neq 0,$ we can find a point satisfying $v-w=1$ and hence a point in $M.$ Thus $M$ is identified with a subset of $P\cN.$''\end{quotation}
This is clear. If $(x,v,w)$ is in $\cN$ and $v-w\neq 0,$ then $\frac{x}{v-w}$ is in $\cN\cap\cZ.$
\begin{quotation} ``The points in $P\cN$ not in $\tau(M)$ corresponds to the generators of $P\cN$ with $v-w=0.$''\end{quotation}
This is evident from the definition. Now there comes an unclear paragraph with an erroneous conclusion:
\begin{quotation}
``This is the intersection of $N$ with a null hyperplane through the origin. All such hyperplanes are equivalent under $O(4,2)$ so to see what these extra points represent, we consider the null hyperplane $v+w=0.$ From Eq. (\ref{eq:tau}) we see that the points of $M$ corresponding to generators of $\cN$ which lie in this hyperplane are just the null cone of the origin. Thus $P\cN$ consists of $\ct(M)$ with an extra cone at infinity.''\end{quotation}
It is rather hard to follow this fuzzy reasoning, therefore we will study the structure of the ``extra part'' directly from the definition. The extra part is the projection by $\pi$ of those points in $\cN$ for which $v-w=0.$ Now the following two cases must be considered separately: either $v=w=0$ or $v=w\neq 0.$ Let $\cN_c=\{Z\in \cN:\, v=w\neq 0,$ and $\cN_s=\{Z\in\cN:\, v=w=0\}.$ Each element of $\pi(\cN_c)$ has a unique representative $Z'=(x',v',w')$ in $\cN$ with $v'=w'=1.$ Since $Q(Z')=0,$ we have $q(x')=0.$ Therefore $\pi(\cN_c)$ has the structure of the null cone at zero in $M.$ But there is also the second part, $\pi(\cN_s).$ If $Z=(\bx,t ,0,0)$ is in $\cN_s$, then $t\neq 0,$ otherwise, because of $Q(Z)=q(x)=0$ we would have $\bx=0$. Therefore each $Z=(\bx,t,0,0)$ in $\cN_s$ has a unique representative with $t=1.$ From $q(x)=0$ it follows then that $\bx^2=1.$ It follows that $\pi(\cN_s)$ has the structure of the 2-sphere. This part is missing in the conclusion of   \cite{tod}. One of the possible reasons for this omission can be a possibly misleading statement in Penrose and Rindler \cite[p. 303]{penrin}, where we can read
\begin{quotation}
``... and the remainder of the intersection of the $4$--plane with ${\tilde M}$ is $\cJ$ (the identified surfaces $\cJ^+,\cJ^-$ of the previous construction).''
\end{quotation}
The point is that in $\cJ$ of Penrose and Rindler one has to first identify the two 2-spheres, one of $\cJ^+$ and one of $\cJ^-,$ though with opposite orientations - see the next subsection. This lack of precision in \cite{penrin} may have confused the authors of \cite{tod,schmidt}.
\subsection{The 2-sphere missed by Akivis and Goldberg\label{sec:akivis}}
A similar inadvertency takes place in a monograph on conformal geometry by M. A. Akivis and V. V. Goldberg \cite{akivis}. In the introductory chapter the authors analyze the Euclidean case. They start with the equation of a hypersphere in the conformal space $C^n,$ which is just $E^{n,0}$ endowed with an Euclidean scalar product defined up to a non--zero multiplicative constant. The equation, in {\it polyspherical coordinates\,} $s^0,s^i,s^{n+1},$ reads:
$ s^0\sum_{i=1}^n(x^i)^2+2\sum_{i=1}^n s^ix^i+2s^{n+1}=0.$
When $s^0\neq 0,$ this can be put in the form:
$$ \sum_{i=1}^n(x^i-a^i)^2=r^2,\quad\mbox{where}\quad a^i=-\frac{s^i}{s^0},\quad r^2=\frac{1}{(s^0)^2}\left(\sum_{i=1}^n(s^i)^2-2s^0s^{n+1}\right).$$
In order to describe a hypersphere of zero radius (centered at $a^i$) we must have
$ (X,X):=\sum_{i=1}^n (s^i)^2-2s^0s^{n+1}=0,$
which is just the equation (\ref{eq:qzz}) of the null cone $\cN$ in $E^{n+1,1}$ with $s^0=\frac{1}{2}(w-v),\, s^{n+1}=(w+v),$ adapted to the Euclidean signature. Hyperspheres of zero radius correspond to the points (their centers) in $C^n.$ The remaining set of non--zero solutions of Eq. (\ref{eq:qzz}) is the line $s^0=0,$ $x^i=0,\,1,\leq i\leq n,$ $s^{n+1}\neq 0$ - the point at infinity.

The same strategy is then used in Chapter 4.1 in the pseudo-Euclidean case. With slight changes of the notation the authors state \cite[p. 127]{akivis} that \begin{quotation}
``... after compactification the tangent space $T_x(M)$ is enlarged by the point at infinity $y$ with coordinates $(0,0,...,0,1)$ and by the isotropic cone $C_x,$ with vertex at this point $y$ whose equation is the same as the equation of the cone $C_x,$ namely $ g_{ij}x^ix^j=0.$'' \end{quotation}
There is a subtle inadvertency there. The change of notation is not important, so let us use the same notation as in the Euclidean case. When $s^0\neq 0$ we have the same situation as in the Euclidean case, except that the ``hypersphere of zero radius'' becomes now a cone (light cone in the Minkowski case). It remains to consider the case of $s^0=0$. Here we have two possibilities: either $s^{n+1}=0$ or $s^{n+1}\neq 0.$ If $s^{n+1}=0,$ then, necessarily, the $n$-vector $(s^i)\neq 0,$ and $g_{ij}s^is^j=0.$ But then, we should consider the set of lines and not the set of points. \\
For example in the case of Minkowski space we find that the set of lines is the quadric ($S^2$), and {\bf not} the ``isotropic cone'', as falsely stated in \cite{akivis}. On the other hand, if $s^{n+1}\neq 0,$ the we can choose $s^{n+1}=1.$ In this case no freedom of choosing the scale remains and we get $g_{ij}s^is^j=0$ - the isotropic cone, including its origin.

Another mistake takes place during the discussion of the conformal inversion in \cite[p. 15-16]{akivis}. The authors state that
\begin{quotation}``In the pseudo-Euclidean space $\BR^n_q,$ the inversion in a hypersphere $S$ with center at a point $A$ is defined exactly in the same manner as it was defined in the Euclidean space $\BR^n$ (...). However, in contrast to the space $\BR^n,$ under an inversion in the space $\BR^n_q$ not only does the center $a$ of the hypersphere $S$ not have an image but also points of the isotropic cone $C_x$ with vertex at the point $a$ does not have images. To include these points in the domain of the mapping defined by the inversion in $\BR^n_q,$ we enlarge the space $\BR^n_q$ not only by the point at infinity, $\infty,$ corresponding to the point $a$ but also by the isotropic cone $C_\infty$ with the vertex at this point. The manifold obtained as the result of this enlargement is denoted by $C^n_q:$
$$ C^n_q=\BR^n_q\;\bigcup \;\{C_\infty\}$$
and is called a {\it pseudoconformal sphere of index $q.$}
(...)
Just like conformal space $C^n,$ the pseudoconformal space $C^n_q$ is homogeneous.''
\end{quotation}
Adding the image of the isotropic cone under inversion does not result in the homogeneous space. In section \ref{sec:inv} we show that the conformal inversion with respect to the isotropic cone $C_0$ centered at the origin $0\in M$ is implemented by the map $(x,v,w)\mapsto (x,-v,w).$ Using the embedding $\tau: M\rightarrow E^{4,1}$ given by Eq. \ref{eq:tau} we find that the image of $C_0$ consists of vectors of the form $(x,\frac12,-\frac12)$ and therefore $\tau(C_0)$ consists of vectors of the form
$(x,-\frac12,-\frac12).$ Let now $x$ be a nonzero vector in $C_0),$ and let $a$ be a vector satisfying $x\cdot a=\frac12.$ The action of the translation group is given by Eq. (\ref{eq:transl}). It is clear that after translation by $a$ the point $(x,-\frac12,-\frac12)$ is mapped  to
$(x,0,0),$ which is {\bf not} in the image of $C_0$ under inversion. Therefore the statement in \cite{akivis} that adding just the image of $C_0$ under inversion gives a homogeneous space is erroneous. It is necessary to add the missing sphere.

A similar misleading statement can be found in a paper by N. M. Nikolov and I. T. Todorov in \cite{todorov}, where the authors state that ``The points at infinity in $\bar{M}$ form a $D-1$ dimensional cone with tip at $p_\infty,$ quoting Penrose \cite{penrose2}, and then state that  ``... the Weyl inversion ... interchanges the light cone at the origin with the light cone at infinity''.\footnote{In a private exchange one of the authors (N.M.N) explained to me that the precise statement should read: ``The Weyl inversion ... interchanges the {\bf compact} light cone at the origin with the compact light cone at infinity, where the compact light ``cone'' with a tip at $p$ is defined as $\{q\in\tilde{M} : p \mbox{ and } q \mbox{ are mutually isotropic }\}.$ '' These concepts have been described in  \cite[Appendix A,C]{nikolov1} and \cite{nikolov2}, and will be developed in their future paper. }

\section{From Einstein's static universe to $P\cN$\label{sec:pn}}
The group $U(1)$ can be identified withe the group of complex numbers $z\in\BC$ with $|z|=1,$ and the group $SU(2)$ can be thought of as the group of unit quaternions $\{q=v+x{\bf i}+y{\bf j}+z{\bf k}\in \BH: \vert q\vert^2=v^2+\bx^2=1\}.$ Let $E^{4,1}$ denote $\BR^5,$ with coordinates $(\bx,v,\psi),$ and endowed with the quadratic form $q_5(\bx ,v,\psi )=\bx^2+v^2-\psi^2.$ Writing $z=e^{i\psi },$ we can then represent the group $U(1)\times SU(2)$ (topologically $S^1\times S^3$) as the cylinder $\mk$ in $E^{4,1}$:
$$ \mk=\{(\bx,v,\psi )\}: \bx^2+v^2=1,\, \psi \in[-\pi,\pi)\}.$$
\begin{lemma}
With $E^{4,2}$ endowed with the coordinates $Z=(\bX,T,V,W),$ as in the previous section (but we will use capital letters here) let $\lambda: E^{4,1}\rightarrow E^{4,2},$ be the map
\be \lambda:\,(\bx,v,\psi)\mapsto(\bX,T,V,W)=(\bx,\sin(\psi),v,-\cos(\psi)).
\label{eq:txvw}
\ee
Then $\pi\circ\lambda$ restricted to $\mk,$ is $2:1$ and surjective: $\pi\circ\lambda(\mk)=P\cN.$ Given any two points $(\bx_1,v_1,{\psi }_1)$ and $(\bx_2,v_2,{\psi }_2)$ in $\mk,$ we have $\pi\circ\lambda(\bx_1,v_1,{\psi }_1)=\pi\circ\lambda(\bx_2,v_2,{\psi }_2)$ if and only if the following conditions (i-iii)
hold
$$
  {\rm ( i)}\, |{\psi }_2-{\psi }_1|=\pi,\quad
  {\rm (ii)}\, \bx_2=-\bx_1,\,{\rm and}\quad {\rm (iii)}\,v_2=-v_1.
$$
\label{lemma:lemma2}
\end{lemma}
\begin{proof}The proof is evident after noticing that $Q(Z)=\bX^2-T^2+V^2-W^2=0$ can be written as $\bX^2+V^2=T^2+W^2,$ and, if $Z\neq 0,$ then $V^2+\bX^2>0.$ Therefore on each generator line of $\cN$ there are exactly two points $Z,-Z,$ with $V^2+\bX^2=1.$ \end{proof}
\vskip0.5cm
In order to be able to represent $P\cN$ graphically, on a plane, let us introduce the map $\rho: \mk\rightarrow [0,\pi]\times [-\pi,\pi]\subset \BR^2$ given by $$\rho: (\bx,v,\psi )\mapsto (\xi=\arccos (v) ,\psi ).$$
\begin{figure}[h!]
\begin{center}
    \leavevmode
      \scalebox{.80}{\includegraphics[width=6cm, keepaspectratio=true]{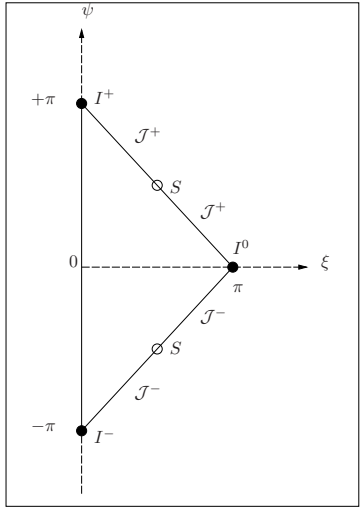}}
\end{center}
  \caption{The ``Penrose diagram'' of Minkowski space.}
\label{fig:fig1}\end{figure}
In Figure \ref{fig:fig1} the resulting ``Penrose diagram'' is shown, using the notation as in \cite[p. 919]{MTW}, but with two distinguished points denoted as $S.$ In this realization they represent one and the same 2-sphere - they need to be identified. The region inside the triangle with vertices at $(0,-\pi),(0,+\pi),(\pi,0)$ corresponds to the points in the Minkowski space. In order to understand this correspondence, let us first notice that owing to the equation $v^2+\bx^2=1,$ we have the following relations:
$$
\bX=\bx,\,
T=\sin(\psi),\,
V=\cos(\xi),\,
W=-\cos(\psi),\,
|\bX|=\sin(\xi).$$
When $V-W\neq 0$, we get the corresponding point in Minkowski space with coordinates $(\br ,t)$ given by the formulae:
\begin{eqnarray*}
\br\quad=\quad\frac{\bX}{V-W}&\;=\;&\frac{\bx}{\cos(\psi)+\cos(\xi)}\\
t\quad=\quad\frac{T}{V-W}&\;=\;&\frac{\sin(\psi)}{\cos(\psi)+\cos(\xi)}\\
r\quad=\quad|\br|\;&=\;&\frac{\sin(\xi)}{\cos(\psi)+\cos(\xi)}.
\end{eqnarray*}
Now, by elementary trigonometric identities we have that:
$$ \tan\left(\frac{\psi+\xi}{2}\right)\quad=\quad\frac{\sin(\psi)+\sin(\xi)}{\cos(\psi)+\cos(\xi)},$$
$$ \tan\left(\frac{\psi-\xi}{2}\right)\quad=\quad\frac{\sin(\psi)-\sin(\xi)}{\cos(\psi)+\cos(\xi)}.$$
It follows that
$$
t+r=\tan\left(\frac{\psi+\xi}{2}\right),\quad
t-r=\tan\left(\frac{\psi-\xi}{2}\right),
$$
which are exactly the equations in \cite[p. 919]{MTW}, and in \cite[p. 121]{HE} (with our $\psi,\xi$ corresponding to their $t^{'},r^{'}$ resp.). Each point in the interior of the triangle represents a 2-sphere at time $t$ and radius $r>0$ centered at the origin of $\bx$-axes. Each points on the open interval $\xi=0, |\psi|<\pi$ represents the origin $(t=0,\br=\bf{0})$ of the Minkowski coordinate system. The points $I^{-}$ and $I^{+},$ with$\xi=0,\psi=\pm\pi$ both correspond to $V=1,W=1,T=0,\bX=0$ - a single point in the compactified Minkowski space, the apex of the null cone $\cN_c$ at infinity. Each point of the open intervals $\mathcal{J}^{\pm}$ corresponds to a 2-sphere $V=W\neq 0,T\neq 0,\bX^2=T^2.$ These 2-spheres build $\cN_c$ except of its apex $I^{+}=I^{-}.$ The point $I^0$ represent the same point of the compactified Minkowski space as $I^{\pm}.$ What is misleading in all the standard literature describing the conformal infinity is the neglecting the fact that there are two exceptional points of the diagram, denoted here as $I^1,$ and corresponding to the parameter values $\xi=\pi/2, \psi=\pm\pi/2.$ These two points correspond to $V=W=0,T=\pm 1,\bX^2=1$ which is the sphere $\cN_s$ discussed at the end of the previous section. These two exceptional points should be identified in order to give the complete representation of the conformal infinity - compare the discussion of these issues in the papers of Roger Penrose \cite{penrose1, penrose2}.
\section{Action of the inhomogeneous Lorentz group (Poincar\'e group) $ISO(3,1)$ \label{sec:il}}
\subsection{Action of $SO(3,1)$}
The homogeneous Lorentz group $SO(3,1)$ maps the conformal infinity into itself. It is thus of interest to analyze this action in some details. We will show that there are two invariant submanifolds for this action, one consisting of a point, and one being the 2-sphere $\cN_s.$ To this end will use the results of W. R\"{u}hl \cite{ruhl}. According to \cite{ruhl}, his Eqs. (2.18), (2.19), the homogeneous Lorentz group is represented by $SU(2,2)$ matrices $\left(\begin{smallmatrix}A&B\\C&D\end{smallmatrix}\right)$ of the form
$
A=D=\frac{1}{2}(R+(R^*)^{-1})\quad {\rm and}\quad
B=C=\frac{1}{2}(-R+(R^*)^{-1})
$
where $\det(R)=1$ and $^*$ denotes Hermitian conjugation.
We need to consider two cases: when $R$ is unitary (pure rotations), and when $R$ is Hermitian (pure boosts). In the case of pure rotations we have $R={R^*}^{-1},$ Therefore, in this case, $A=D=R, B=C=0,$ and the fractional linear action of $SU(2,2)$ on $U(2)$ becomes
$ U^{'}=RUR^{-1}.$
It is clear that the point at infinity corresponding to $U=E$ is invariant. Also the spectrum of $U$ is an invariant of this transformation, therefore the 2-sphere $\cN_s$ corresponding to $U$ with eigenvalues $\pm1$ is mapped into itself.\\
Now consider the boosts, with $R=R^*.$ Denote $R_+=R+(R^*)^{-1},$ $R_{-}=-R+(R^*)^{-1}.$ The fractional linear transformation corresponding to the boosts are then of the form:
\be U^{'}=(R_+U+R_{-})(R_{-}U+R_+)^{-1}.\label{eq:up}\ee
Evidently the point $U=E$ is left invariant. Consider now the 2-sphere $\cN_s$ corresponding to the unitary operators $U$ with eigenvalues $\pm1.$ These points are characterized by the property $U^2=E.$ Therefore we can rewrite the Eq. (\ref{eq:up}) as
$ (R_+U+R_{-})((R_{-}+R_+U)U)^{-1}=ZUZ^{-1},$
where $Z=R_+U+R_{-}.$ It follows that then also $(U^{'})^2=E,$ therefore the Lorentz boosts map the 2-sphere $\cN_s$ onto itself. Thus $\cN_s$ is an $O(3,1)$--invariant submanifold of the conformal infinity.
\subsection{Action of the translations\label{sec:translations}}
Consider the translation by a four--vector $a\in M.$ Using Clifford algebra methods and the formula for the translations in \cite[p. 87]{angles} it is easy to calculate the effect of the translation in terms of variables $(x,v,w)$ of section \ref{sec:ref2}:
\begin{eqnarray}
x^{'}&=&x-(v-w)a\\
v^{'}&=&v+(x\cdot a)-\frac{a^2}{2}(v-w)\\
w^{'}&=&w+(x\cdot a)-\frac{a^2}{2}(v-w)
\label{eq:transl}\end{eqnarray}
At the conformal infinity we have $v=w$, therefore $x^{'}=x,$ but, for $x\neq 0,$  the coordinates $v$ and $w$ change. If $v=w=0,$ then, after the generic translation, $v^{'}=w^{'}\neq 0.$ The coordinate description of the 2-sphere $\cN_s,$ which is the common part of $\mathcal{J}^+$ and $\mathcal{J}^{-}$ changes. What is invariant is the set $\mathcal{J}^+\cup \mathcal{J}^{-},$ and the fact that $\mathcal{J}^+$ and $\mathcal{J}^{-}$ have a common 2-sphere.
\subsection{Transitivity of $ISO(3,1)$ on the conformal infinity}
Let $\cJ$ denote the conformal infinity, {\em minus the singular point\,} $I_0=I_+=I_-.$ It is easy to see that action of $ISO(3,1)$ on the is transitive. $\cJ$ has the topology of a cylinder $\BR\times S^2.$ The group of translations acts along the $\BR,$ while $SO(3,1)$ acts transitively on $S^2$ in a standard way - Lorentz transformations act on directions of light rays through the origin of the Minkowski space. It follows that any splitting of $\cJ$ into $\cJ^+$ and $\cJ^-$ is not translation invariant and not intrinsic. The article of Roger Penrose \cite{penrosesst} is extremely unclear in this respect.
Penrose mentions for instance that ``There is another version of compactified Minkowski space in which the future boundary hypersurface is identified with the past'', and quotes his earlier paper \cite{penrose1967}, as well as the classic one by Kuiper \cite{kuiper}, but he does not bother to define precisely what would be the alternative for the projective model. The same lack of clarity concerns the discussion in \cite{HE} and \cite{MTW}. B.G. Schmidt, in an apparently mathematically precise paper \cite{schmidt} proves a Theorem stating that {\em The conformal boundary of Minkowski space is\,} $\cJ^+\cup \cJ^-\cup I^+\cup I^-\cup I^0,$ without ever bothering to define the sets on the right hand side of his statement.

In   \cite[p. 178]{penrose1965} Penrose writes:
\begin{quotation}``There is one property of ${\mathcal R},$ however, which seems undesirable when these ideas are applied to interacting fields, or curved space--times. This is the fact that the `future infinity' turns out to have been identified with the `past infinity' in the definition of ${\mathcal R}.$ To avoid this feature it will be desirable effectively to `cut' this manifold along the hypersurface $\cJ$ and to consider instead the resulting manifold with boundary. This boundary consists essentially of two copies of $\cJ,$ one in `future' which will be called $\cJ^+$ and one in the `past to be called $\cJ^-$ ....''\end{quotation}
Nowhere a precise definition of $\cJ^+$ and $\cJ^-$ is given. We are not told how the Poincar\'{e} group acts on these `boundaries'. Also the authors of recent papers like, for instance \cite{zengi}, when asked about the definition of $\cJ^+$ and $\cJ^-$ for Minkowski space, refer to Penrose \cite{penrose1965} or \cite{geroch}. In fact Geroch does not define $\cJ^+$ and $\cJ^-$ for the Minkowski space. He considers Schwarzschild space--time with the topology $S^2\times \BR^2,$ proposes some coordinate-dependent constructions and does not really discuss global symmetries.
\section{Light trapped at infinity\label{sec:geo}}
The aim of this section is to demonstrate that a light ray can be trapped in the conformal infinity and circulate there forever - unless disturbed by some quantum effect. It is well known (cf. e.g.,   \cite{wk} for a clear and self--contained exposition) that null geodesics are described by two-dimensional totally isotropic subspaces of $E^{4,2}.$ Using the coordinates $(x,v,w)$ as in Sec. \ref{sec:ref2}, let $x_0$ be a fixed non--zero null vector in $E^{3,1},$ and let $n_1$ and $n_2$ be the vectors in $E^{4,2}$ defined by
$
n_1=(x_0,0,0)\quad {\rm and}\quad
n_2 = (0,1,1).$
Then the two-dimensional (real) plane spanned by $n_1$ and $n_2$ is totally isotropic - therefore it is describing a null geodesic in the compactified Mink\-owski space. A general vector in this plane is of the form $\alpha n_1+\beta n_2=(\alpha x_0,\beta,\beta ), $ therefore it is completely contained in the conformal infinity that consists of null vectors $(x,v,w)$ with $v=w.$ We can completely parameterize our null geodesic by a parameter $\tau\in[0,\pi]$ by choosing the representatives of its points in the form
\be(\cos(\tau ) x_0,\sin(\tau ),\sin(\tau )).\label{eq:light}\ee
For $\tau=0$ the geodesic is on the 2-sphere $\cN_s,$ for $\tau=\pi/2$ it reaches the exceptional point $I^+=I^-=I^0,$ then it circulates further towards the 2-sphere $\cN_s.$  Notice that for $\tau=\pi$ we get the point $(-x_0,0,0)$ which projects onto the same point of $P\cN$ as $(x_0,0,0).$ Replacing $x_0$ by $\lambda x_0,\,\lambda\in \BR$ does not change the plane spanned by $n_1,n_2,$ therefore in this way we get a family of null geodesics, all trapped in the conformal infinity. We can always choose a representative of $x_0$ of the form $(\br ,1),$ $\br^2=1,$ so that we have a trapped null geodesic for every point of the unit sphere in $\BR^3.$
\subsection{Conformal inversion\label{sec:inv}}
Consider the following linear map $R$ of $E^{4,2}:$ $R:(x,v,w)\mapsto (x,-v,w).$ It is clear that $R\in O(4,2)$ (though not in $SO(4,2)$). It is instructive to see that $R$ implements the conformal inversion $x\mapsto x/x^2$ of the Minkowski space. To this end let $x$ be a point in the Minkowski space $M$ and let, writing $x^2$ for $q(x),$ $\ct(x)=\left(x,(1-x^2)/2,-(1+x^2)/2\right)$
be its image in $E^{4,2}$ as in Eq. (\ref{eq:tau}).\footnote{This is the standard map, discussed in a general signature for instance in \cite[p. 92]{angles}.}  We apply the inversion $R$ to obtain $\left(x,-(1-x^2)/2,-(1+x^2)/2\right)$
and represent it as an image of a new point $x'.$ Therefore we should have
\be \left(x',-\frac{1}{2}(1-x'^2),-\frac{1}{2}(1+x'^2)\right)=\lambda \left(x,\frac{1}{2}(1-x^2),-\frac{1}{2}(1+x^2)\right).\label{eq:cil}\ee Now, from $x'=\lambda x$ it follows that $x'^2=\lambda^2 x^2.$ Substituting this value of $x'^2$ into the two other equations and adding them we get $\lambda=1/x^2,$ therefore
$ x'=\frac{x}{x^2},$
which is the well known conformal inversion in Minkowski space. The formula (\ref{eq:cil}) becomes then an identity.\footnote{It is evident that this formula makes sense only when a length scale is chosen. This can be a Planck length, a cosmic scale length or some other length scale. The formula is singular on the light cone, but this apparent singularity is a coordinate effect.}

Let us now apply the conformal inversion $R$ to the light rays circulating at infinity, given by the formula (\ref{eq:light}). We obtain the family
$$\left(x_0,-\sin(\tau ),\sin(\tau )\right)=-2\sin(\tau )\left(x(\tau),\frac{1}{2}(1-x(\tau)^2),-\frac{1}{2}(1+x(\tau)^2\right),$$
where $x(\tau)=-\frac{1}{2}\cot (\tau)x_0.$ This is nothing else but a family of light rays through the origin of the Minkowski space in the directions of null vectors $x_0.$ The parameter $\tau$ is, of course, not an affine parameter of these null geodesics.
\subsection{The signature of the metric at infinity}
Let $H_1$ be the affine hyperplane in $E^{4,2}$ parameterized by the coordinates $(\br ,t,v,w),$ defined by the condition $t=1.$ Then $H_1$ is transversal with respect to the null cone $\cN, $ therefore, by Theorem $3$ of   \cite{wk} it induces the unique conformal structure on $\pi(H_1\cap\cN).$ The intersection $H_1\cap\cN$ is described by the equation $\br^2-1+v^2-w^2=0.$ Taking a trajectory there, by differentiation,  we get for the tangent vector $({\dot \br},{\dot v},{\dot w})$ the equation $\dot{\br}+\dot{v}-\dot{w}=0.$ Notice that at the points corresponding to the conformal infinity we have $v=w.$ Taking a trajectory with $v=w=\mbox{const}$ we get a trajectory on the 2-sphere. The signature there is $(2,0).$ On the other hand, taking a trajectory with $\br$ constant we obtain a tangent vector of the form $({\bf 0},0,{\dot v},{\dot w})$ - a null vector in $E^{4,2}.$ It follows that the metric induced on conformal infinity is degenerate and has as its standard form $\diag(1,1,0)$
\subsection{A pictorial representation of the infinity}In order to get an idea about the manifold structure of the conformal infinity and to obtain its pictorial representation, it is convenient to use the formulas from Lemma \ref{lemma:lemma2}. At the conformal boundary we have $v=w,$ thus $v=\cos(\psi),$ and since $v^2+\bx^2=1,$ we get $\bx^2=\sin^2(\psi).$ Furthermore, because $(\bx,t ,v,w)$ and $(-\bx,-t ,-v,-w)$ describe the same point of $P\cN,$ it is enough to consider $\psi\in [0,\pi].$ The whole conformal infinity is then described by one equation:
$$ \bx^2=\sin^2 (\psi ),\quad \psi\in[0,\pi],$$
where $(\bx,\psi=0)$ and $(\bx,\psi=\pi )$ describe the same point. This is nothing else but a squeezed torus. Replacing the spheres $S^2$ by circles $S^1$ we get the graphic representation as shown in Fig. \ref{fig:fig4}. Topology itself is represented by a double cone with two vertices identified, as in Fig. \ref{fig:fig3}. This picture must not be confused with a similarly looking picture taken from   \cite[p. 178]{beem}, which we reproduce here in Fig. \ref{fig:fig3a}.
\subsection{The double covering\label{sec:dc}}
It is possible to repeat the constructions of Sects \ref{sec:ref2} and \ref{sec:pn}, but replacing the equivalence relation $Z\sim Z'$ by a stronger one: we identify two vectors $Z$ and $Z'$ in $E^{4,2}$ is and only if $Z'=\lambda Z,\, \lambda>0.$ The manifold resulting by taking the quotient of $\cN$ by this new equivalence relation will be denoted by ${\hat P}\cN.$ Instead of one map $\tau$ as in Eq. (\ref{eq:tau} we define now two maps:
\be \ct_+(x)=(x,\frac{1}{2}(1-q(x)),-\frac{1}{2}(1+q(x))).\label{eq:taup}\ee
\be \ct_-(x)=(x,-\frac{1}{2}(1-q(x)),\frac{1}{2}(1+q(x))).\label{eq:taum}\ee
Similarly we define
\be \cZ_\pm=\{Z\in E^{4,2}:\, v-w=\pm 1\},\label{eq:zpm}\ee
and then show that
\begin{lemma} The image $\ct_\pm (E^{3,1})$ in $E^{4,2}$ coincides with the intersection $\cN\cap\cZ_\pm$ of $\cN$ with $\cZ_\pm.$
\end{lemma}
The manifold ${\hat P}\cN$ contains now two copies of Minkowski space, we may call them $M_{+}$ and $M_{-},$ joined by a common boundary.
\vskip10pt
\begin{figure}[h!]
\begin{center}
    \leavevmode
      \includegraphics[width=5cm, keepaspectratio=true]{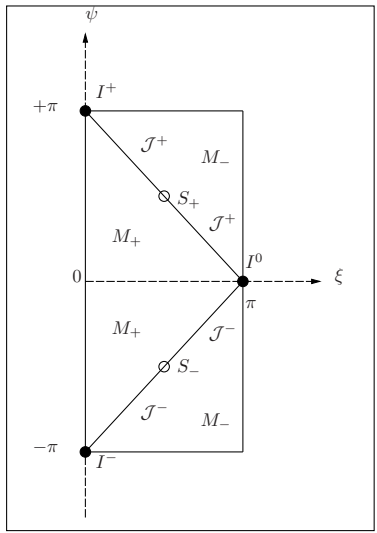}
\end{center}
  \caption{The second version of the ``Penrose diagram'' of Minkowski space.}
\label{fig:fig2}\end{figure}
In Figure \ref{fig:fig2} the corresponding Penrose diagram is shown, this time we have two different 2-spheres $S^+$ and $S^{-}.$ There are two copies of Minkowski space, $M_{+}$ and $M_{-},$ separated by the boundary. The horizontal lines at $\psi=+\pi$ and $\psi=-\pi$ should be identified. The corresponding pictorial representation of the infinity is shown in Figures \ref{fig:fig3}, \ref{fig:fig4}.

\section{Geometry of oriented twistors \label{sec:ot}}
In this section we present a slightly modified version of the reasoning of Kopczy\'{n}\-ski and Woronowicz in \cite[section III]{wk}\footnote{Our numbering conventions differ slightly from those used in \cite{wk}. We use Roman letters $e,x,y,v,w,$ etc. to denote the elements of the algebra. A different approach, using pure Clifford algebra methods and dealing with the case of non--oriented twistors, is discussed by Crumeyrolle \cite[Ch. 12. Twsitors]{crumeyrolle}}. In particular will take into account the orientation, and also we will change the notation a little bit by introducing the Hodge $\star$ operator. Otherwise, in this section we will follow the notation of the   \cite{wk} - that may differ from the notation in other parts of this paper.
To start with: as it will be explicitly shown below in section \ref{sec:cliff}, twistors are spinors for the conformal group\footnote{In his Afterward to ``Such Silver Currents. The Story of William and Lucy Clifford 1845-1929'' \cite[p. 182]{chisholm} Roger Penrose wrote: {\em Twistors may be regarded as spinors for six dimensions; yet they refer directly to the four dimensions of space--time.} In ``The Road to Reality'' \cite[Ch. 33.4]{penrose-r}  Penrose writes {\em How do twistors fit in with all this? The shortest — but hardly the most transparent — way to describe a (Minkowski-space) twistor is to say that it is a reduced spinor (or half spinor) for O(2, 4).}  }. But, for our present purpose, in order to analyze the twistor geometry no knowledge of spinors is needed. We will make this section self--contained - to a large extent. Nevertheless it may be useful to recall the fact that the spinor space for the conformal group is the space of an irreducible representation of the even Clifford algebra $Cl^+_{4,2}$, the dimension of this space over $\BC$ being $2^{\frac{r+s}{2}-1}=4,$ which is the same as the dimension of $H_{2,2}.$
\subsection{The exterior algebra $\bigwedge\, V $ and Hodge duality operator}
Let $ V $ be a complex vector space of finite dimension $n.$  We denote by $\bigwedge  V =\bigoplus_{k=0}^n \bigwedge^k\,  V $ the exterior algebra of\; $ V $ thought of as a consisting of antisymmetric tensors endowed with the wedge product\footnote{For more information about exterior (Grassmann) algebras see e.g.,   \cite[Ch. 5]{greub}.}:
$$ v_1\wedge ...\wedge v_k=\sum_\sigma (-1)^\sigma v_{\sigma(1)}\otimes ...\otimes v_{\sigma(k)}.$$
Assume that $V$ is endowed with a pseudo--hermitian form $(x|y)$ of signature $(p,q).$ The standard example is the space $\BC^n=\BC^p\oplus\BC^q$ with
  $$ (x|y) =\sum_{i=1}^p x^i{\bar y}^i-\sum_{j=1}^q x^j{\bar y}^j.$$
  We endow $\bigwedge  V $ with a natural pseudo--hermitian form defined by:
\be ( v_1\wedge ... \wedge v_k|w_1\wedge ...\wedge w_k) =\det\left( (v_i|w_j)\right).\label{eq:scf}\ee
\begin{remark} Notice that there exist, in the literature, two different conventions of defining the exterior product. While most authors seem to agree on the definition of the alternating operator:
$$ \mbox{Alt}\,(v_1\otimes ... v_k)=\frac{1}{k!}\sum_\sigma (-1)^\sigma v_\sigma(1)\otimes ...v_\sigma (k),$$
the exterior product of a $k$--vector $v$ and $l$--vector $w$ can be defined by the formula:
$$ v\wedge w =\left(\frac{(k+l)!}{k!l!}\right)^\epsilon \mbox{Alt}\,(v\otimes w),$$
where $\epsilon=0$ or $\epsilon=1.$ We choose $\epsilon=1.$\\
There are also two different convention of extending the scalar product from $ V $ to $\bigwedge\, V .$ Some authors (especially physicists, when discussing the second quantization of Fermions) endow $\bigwedge\, V $ with the restriction of the natural scalar product defined on the tensor product. For $k$--vectors this gives $k!$ times our scalar product.
\end{remark}
Given $x\in\bigwedge^p V$ we have the coordinate representation of $x$ in a basis $\{e_i\}$ of $V$:$$ x=\frac{1}{p!}\,x^{i_1...i_p}\,e_{i_1}\wedge ...\wedge e_{i_p}.$$ The wedge product is then given by the formula:
$$ (x\wedge y)^{i_1...i_{p+q}}=\frac{1}{p!q!}\;\delta^{i_1...\phantom{j_pj_{p+1}}...i_{p+q}}_{j_1...j_pj_{p+1}...j_{p+q}}\;x^{j_1...j_p}\;y^{j_{p+1}...j_{p+q}},$$
where $\delta^{a...b}_{c...d}$ is the (generalized) Kronecker delta symbol. We also have the coordinate representation:
\be (x|y) = \frac{1}{p!}\;G_{i_1j_1}..G_{i_p j_p}x^{i_1...i_p}\;\overline{y^{j_1...j_p}},\label{eq:xy}\ee
where $G_{ij}=\langle e_i,e_j\rangle .$\vskip0.5cm
Let now $\{e_i\}$ be an orthonormal basis for $ V $ with $(e_i|e_i)=+1$ for $i=1,... ,p,$ and $=-1$ for $i=p+1,...,p+q,$ and let $e=e_1\wedge ...\wedge e_n.$ Then $(e|e) =(-1)^q.$ Let $e\in\bigwedge^n V$ be a unit $n$--vector. We call $e$ an {\em orientation\,} of $ V.$ An orthonormal basis $\{e_i\}$ will be called {\em oriented\,} if $e_1\wedge ... \wedge e_n=e.$ Any two oriented bases are then related by a unique transformation from the group $SU(r,s).$\vskip0.5cm
For each $x\in \bigwedge V $ let $C(x)$ be the linear operator on $\bigwedge V $ defined by
$$ C(x)y=x\wedge y.$$
Clearly, for $x\in\bigwedge^k V,$ we have $C(x):\bigwedge^l V\rightarrow \bigwedge^{k+l} V,$ and
 $v\mapsto C(v),\quad v\in V$ is a linear map from $ V $ to $L(\bigwedge V),$ with \be C(v)C(w)+C(w)C(v)=0,\label{eq:car1}\ee for all $v,w\in V .$ Notice that it follows from the definition that $C(x\wedge y)=C(x)C(y).$

Let $C(x)^*$ be the Hermitian adjoint of $C(x),$ defined by $$(C(x)^* y|z)=(y|C(x)z),\quad y,z\in\bigwedge V.$$ Then, for $x\in \bigwedge^k V,$ $C(x)^* :\bigwedge^l V\rightarrow \bigwedge^{l-k} V,$  the map $x\mapsto C(x)^*$ is anti--linear, and for $v,w\in V$ we have the anti--commutation relations:
\be C(v)C(w)^*+C(w)^* C(v)=(v|w).\label{eq:car2}\ee
Notice that for all $x,y\in\bigwedge V$ we have $C(x\wedge y)^*=C(y)^\star C(x)^*.$
\begin{remark}
The anti--commutation relations (\ref{eq:car1},\ref{eq:car2}) are known as CAR - canonical commutation relations - in our case finite--dimensional and generalized for the case of an indefinite scalar product. If we define $\phi(v)=2^{-\frac12}\left(C(v)+C(v)^*\right),$ then the real linear map $v\mapsto \phi(v)$ is a Clifford map for $V$ considered as a $2n$--dimensional real vector space endowed with the scalar product $\Re\left( (v|w)\right)$ - cf. \cite{baez}\footnote{For a complex number $z=\alpha+i\beta$ we denote $\Re (z)=\alpha,$ $\Im (z)=\beta.$}
\end{remark}
Assuming $V$ oriented with an orientation $e,$ we define the Hodge operator $\star :\bigwedge^k\, V \rightarrow \bigwedge^{n-k}\, V $ as an antilinear map $\star:x\mapsto \star x$ uniquely defined by the formula
\be x\wedge \star y = (x|y)e,\quad x,y\in\bigwedge^k\, V .\label{eq:sto}\ee
It is easy to see that an equivalent definition of the Hodge $\star$ operator is given by:
$$ \star x=C(x)^*\,e.$$
It easily follows from the definition that for $x\in\bigwedge^k V,$ $y\in\bigwedge^{n-k} V$ we have:
\be (x|\star y)=(-1)^{k(n-k)}(y|\star x).\label{eq:star}\ee
A little bit more effort\footnote{Cf. e.g., \cite[p. 167]{curtis}, \cite[p. 118]{fecko}} is required to check that we have
$$ \star\star x=(-1)^{k(n-k)+q}\,x,\quad \forall x\in\bigwedge^k V.$$
\begin{remark}
A $k$--vector $x\neq 0$ is called {\bf decomposable\,} if $x$ is of the form $x=x_1\wedge ..\wedge x_k$ for $x_1,...,x_k\in V.$ If $x$ is decomposable, then also $\star x$ is decomposable. Moreover the $(n-k)$--dimensional subspace corresponding to $\star x$ is the orthogonal complement of the subspace corresponding to $x$ - cf. \cite[Exercise 8, p. 62]{bourbaki1}.
\label{rem:sd}\end{remark}
Another important property involving creation and annihilation operators to the Hodge star operator is \cite[eq. 139]{sto} is\footnote{While only positive definite scalar product is discussed in   \cite{sto}, this particular property can be easily seen to hold also for pseudo--Hermitian spaces.}
$$ \star\, C(x)^*\star^{-1}=C(x)(-1)^{d(x){\hat N}},$$
where $d(x)$ is the grade of $x$ ($d(x)=k$ for $x\in \bigwedge^k V$) and ${\hat N}$ is the number operator - ${\hat N}y=ly$ for $y\in\bigwedge^l V.$

We define a {\em bilinear\,} form
$$\langle x,y\rangle =(x|\star y),\quad x,y\in \bigwedge V.$$
Notice that the following formulas hold:
$$\langle x,y\rangle =(-1)^q \langle y,x\rangle,\quad
\overline{\langle x,y\rangle}=(-1)^{k(n-k)+q}\langle x,y\rangle$$
In an orthonormal basis $e_i$ such that $e=e_1\wedge ... e_n$ we have the explicit expression for the star operator for $x\in\bigwedge^p V:$
\be (\star x)^{i_{p+1}....i_n}=\frac{1}{p!}\,G_{i_1j_1}... G_{i_pj_p}\,\epsilon^{i_1...i_pi_{p+1}...i_n}\,\overline{x^{j_1...j_p}}.\label{eq:starex}\ee
\subsection{The case of signature $(2,2)$}
In this section we specialize to the case of the signature $(2,2)$ that is relevant for our purposes, and has been studied in   \cite{wk}.

Let $G$ be the diagonal matrix $G=\diag (+1,+1,-1,-1).$ Let $H_{2,2}$ be a four--dimensional complex vector space endowed with a pseudo--Hermitian form $(\cdot|\cdot )$ of signature $(2.2).$ A basis $e_i$ of $H_{2,2}$ is said to be orthonormal if $(e_i|e_j)=G_{ij}.$
Any two orthonormal bases are related by a transformation from the group $U(2,2).$ We fix an orientation $e\in\bigwedge^4 H_{2,2}$ and define the Hodge $\star$ duality operator as in previous subsection). Notice that $\bigwedge^2 H_{2,2}$ we have $\star^2=1.$ Let $\Re\bigwedge^2 H_{2,2}$ be the space of self--dual bivectors:
$$ \Re\bigwedge^2 H_{2,2} =\{x\in \bigwedge^2 H_{2,2}:\, x=\star x\}.$$
Then $\Re\bigwedge^2 H_{2,2}$ is a six--dimensional real vector space, and the real--bilinear form $\langle x,y\rangle$ is real--valued and symmetric on $\Re\bigwedge^2 H_{2,2}.$ It can be easily seen (Cf. \cite[Theorem 7]{wk}) that $\Re \bigwedge^2 H_{2,2}$ equipped with the scalar product $\langle x,y\rangle$ is of signature $(4,2).$ It follows that all the constructions of section \ref{sec:dc} apply and in the following we will use the notation of this section. In particular we will the identification $E^{4,2}=\Re\bigwedge^2 H_{2,2}.$

In a complex vector space the concept of an orientation of a subspace is not well defined. In our case, however we can define what is meant by an oriented two--dimensional subspace. Given a k-dimensional subspace $S$ we can associate with it a simple (i.d. decomposable) nonzero $k-vector$ $x,$ unique up to a non-zero complex factor. For $\lambda\neq 0$ $x$ and $\lambda x$ define the same subspace. For $k=2$ we can restrict the freedom of choice by demanding that $x$ should be self--dual: $\star x=x.$ This restricts the freedom of choice to $\lambda$ real - that is either positive or negative. By an ``oriented two-space'' we will thus mean an equivalence class of simple self--dual bivectors, where $x$ and $y$ define the same oriented subspace if and only if $y=\lambda x,\, \lambda>0.$

Consider now the Grassmann manifold of oriented {\em totally isotropic\,} (complex) subspaces of $H_{2,2}.$ We can repeat now, slightly modified, argument of \cite{wk}.\footnote{For an additional information related to this subject, see also \cite{lernerjmp,lernercmp}.}

\begin{thm}
There is a one--to--ne correspondence between the elements of ${\hat P}\cN$ (the double covering of the compactified Minkowski space), and the oriented isotropic subspaces of $H_{2,2}.$
\end{thm}
\begin{proof} If $p\in {\hat P}\cN,$ then there exists a unique up to a multiplication by a positive constant, non--zero element $x$ of $E^{4,2}$ in the equivalence class of $p.$ Since $x$ is a null vector of $E^{4,2},$ and since, as a bivector, it is self--dual, it follows that $x\wedge x=x\wedge\star x=(x|x)e=(x|\star x)e=\langle x,x\rangle e=0.$ Therefore $x$ is decomposable and it represents a two--dimensional subspace $S(q).$ Now, since $x$ is self dual, $x=\star x,$ it follows from the Remark \ref{rem:sd} that $S(q)$ is orthogonal to itself, and thus totally isotropic as a subspace of $H_{2,2}$. Conversely, let $x$ be a self--dual bivector representing an oriented totally isotropic subspace $S$. Then $(x|x)=0$ (since the subspace is totally isotropic), and, since $\star x=x,$ we have $\langle x,x\rangle =0,$ thus $x$ is an isotropic vector of $E^{4,2},$ and therefore determines $p\in {\hat P}N.$
\end{proof}
\subsubsection{Relation to the U(2) compactification}
In section {\ref{sec:ccms} the points of ${\tilde M}$ have been described by unitary operators $U\in U(2),$ while in this Section by rays in the space of self--dual null bivectors in $E^{4,2}.$ It may be of interest to derive an explicit formula connecting these two descriptions.

Let us equip $H_{2,2}$ with an orthonormal  basis $e_1,e_2,e_3,e_4$ and orientation $e=e_1\wedge ... \wedge e_4.$ Then $H_{2,2}$ can be decomposed into $H_{2,2}\approx \BC^2\oplus \BC^2,$ and every vector $x\in H_{2,2}$ can be written as
$x=\left(\begin{smallmatrix}u\\v\end{smallmatrix}\right),\, u,v\in\BC^2.$ It is easy to see that there is a bijection between unitary matrices $U$ in $\BC^2$ and maximal totally isotropic subspaces in $H_{2,2}:$  Every maximal totally isotropic subspace $\cW$ of $H_{2,2}$ is of the form
$ \cW=\{\left(\begin{smallmatrix}Uv\\v\end{smallmatrix}\right):\,u\in \BC^2\},$
where $U$ is uniquely determined by $\cW.$
Conversely, given unitary $U$ the above formula defines a $2$--dimensional maximal totally isotropic subspace $\cW.$ For our purposes it will be convenient to write the unitary operators as $cU,$ where $c$ is in $U(1),$ (i.e., $\{c\in \BC:\, |c|=1\}),$ and $U$ is in $SU(2).$ To each $(c,U)\in U(1)\times SU(2)$ we associate a maximal totally isotropic subspace $\cW(c,U)$ defined by
$ \cW(c,U)=\{\left(\begin{smallmatrix}Uv\\cv\end{smallmatrix}\right):\,v\in \BC^2\}.$
Till now we still have a redundancy, since $(c,U)$ and $(-c,-U)$ define the same subspace. However, this redundancy will soon disappear when we will move from subspaces to oriented bivectors. In order to do this select two basis vectors in $\BC^2:$
$ v_1=\left(\begin{smallmatrix}1\\0\end{smallmatrix}\right),\, v_2=\left(\begin{smallmatrix}0\\1\end{smallmatrix}\right),$
and let $f_i(c,U)\in H_{2,2},\, i=1,2$ be defined by
$ f_i=\left(\begin{smallmatrix}Uv_i\\ cv_i\end{smallmatrix}\right).$
Every matrix $U\in SU(2)$ can be uniquely written in the form
$ U=\left(\begin{smallmatrix}\bar{\alpha}&\beta\\-{\bar\beta}&\alpha\end{smallmatrix}\right),\quad |\alpha|^2+|\beta|^2=1.$
Our vectors $f_i$ can then be written in components as follows:
$ f_1=\left(\begin{smallmatrix}\bar{\alpha}\\-{\bar\beta}\\c\\0\end{smallmatrix}\right),\quad f_2=\left(\begin{smallmatrix}\beta\\\alpha\\0\\c\end{smallmatrix}\right),$
or
$ f_1=\alpha e_1-{\bar\beta}e_2+c\, e_3\quad \mbox{and}\quad f_2=\beta e_1+{\bar\alpha}e_2+c\, e_4.$
To the pair $(c,U)$ we associate the bivector $f_1\wedge f_2,$ easily calculated to be
$ f_1\wedge f_2=e_{12}-c\beta e_{13}+c\bar{\alpha} e_{14}-c\alpha e_{23}-c{\bar\beta}e_{24}+c^2e_{34},$
where $e_{ij}=e_i\wedge e_j.$ It follows by the very construction that $f_1\wedge f_2$ is a null vector in $\bigwedge^2 H_{2,2},$ what can be easily checked, but it is not, in general,  self--dual: $\star (f_1\wedge f_2)\neq f_1\wedge f_2.$ Therefore let us consider bivector $f$ defined by the formula:
$ f=\frac{1}{\sqrt{2}}\left(f_1\wedge f_2+\star (f_1\wedge f_2)\right).$
Now $f(c,U)$ is both null and self--dual.

From the explicit formulas (\ref{eq:xy}), (\ref{eq:starex}) we easily find the following properties of the basis vectors $e_i,\, i=1,\ldots,4$:
$$ 1=(e_{12}|e_{12})=(e_{34}|e_{34})=-(e_{13}|e_{13})=-(e_{14}|e_{14})=-(e_{23}|e_{23})=-(e_{24}|e_{24}),$$
and
$
\star e_{12}=e_{34},$
$\star e_{13}=e_{24},$
$\star e_{14}=-e_{23},$
$\star e_{23}=-e_{14},$
$\star e_{24}=e_{13},$
$\star e_{34}=e_{12}.$
Define the following six $4\times 4$ antisymmetric matrices $\Sigma_i=(\Sigma_i^{AB})$:\footnote{These matrices have been constructed using the fact that $Cl_{4,2}=Cl_{3,1}\otimes Cl_{1,2}=(Cl_{2,0}\otimes Cl_{1,1})\otimes Cl_{1,1},$ constructing this way $8\times 8$ real matrix generators, then finding a unique (up to scale) metric matrix invariant under the Clifford group - of signature $(4,4),$ and also a unique invariant complex structure in $\BR^8,$ the expressing the generators as complex $4\times 4$ matrices, and renumbering them generators.}
$$ \Sigma_1=\left(\begin{smallmatrix}
 0 & 0 & -i & 0 \\
 0 & 0 & 0 & i \\
 i & 0 & 0 & 0 \\
 0 & -i & 0 & 0
\end{smallmatrix}
\right)\, \Sigma_2=\left(
\begin{smallmatrix}
 0 & 0 & -1 & 0 \\
 0 & 0 & 0 & -1 \\
 1 & 0 & 0 & 0 \\
 0 & 1 & 0 & 0
\end{smallmatrix}
\right)\,
\Sigma_2=\left(
\begin{smallmatrix}
 0 & 0 & -1 & 0 \\
 0 & 0 & 0 & -1 \\
 1 & 0 & 0 & 0 \\
 0 & 1 & 0 & 0
\end{smallmatrix}
\right)\,
\Sigma_3=\left(
\begin{smallmatrix}
 0 & 0 & 0 & i \\
 0 & 0 & i & 0 \\
 0 & -i & 0 & 0 \\
 -i & 0 & 0 & 0
\end{smallmatrix}
\right)$$
$$\Sigma_4=\left(
\begin{smallmatrix}
 0 & i & 0 & 0 \\
 -i & 0 & 0 & 0 \\
 0 & 0 & 0 & -i \\
 0 & 0 & i & 0
\end{smallmatrix}
\right)\quad \Sigma_5=\left(
\begin{smallmatrix}
 0 & 0 & 0 & 1 \\
 0 & 0 & -1 & 0 \\
 0 & 1 & 0 & 0 \\
 -1 & 0 & 0 & 0
\end{smallmatrix}
\right)\quad
\Sigma_6=\left(
\begin{smallmatrix}
 0 & 1 & 0 & 0 \\
 -1 & 0 & 0 & 0 \\
 0 & 0 & 0 & 1 \\
 0 & 0 & -1 & 0
\end{smallmatrix}
\right)
$$
\begin{lemma}
The following identities hold:
\be \overline{\Sigma_\alpha^{ij}}=\frac{1}{2}\epsilon^{ijkl}G_{km}G_{ln}\Sigma_\alpha^{mn},\label{eq:sdsigma}\ee
where $\alpha=1,...,6,$ $i,j,k,l,m,n=1,...,4.$
\end{lemma}
\begin{proof} Easily follows by a direct calculation.\footnote{These and some other calculations in this paper have been aided by several different computer algebra systems.}
\end{proof}
It follows from the Eq. (\ref{eq:sdsigma}) that if we define bivectors $E_1,...,E_6$ by the formula
$ E_\alpha=\frac{1}{2\sqrt{2}}\,\Sigma_\alpha^{ij}\,e_i\wedge e_j,$
then $ \star E_\alpha=E_\alpha.$ Moreover, one can verify that we have
$ \langle E_\alpha,E_\beta\rangle=Q_{\alpha\beta},$ where \be Q=\mbox{diag }(1,1,1,-1,1,-1).\label{eq:q}\ee
Explicitly we have:
$$
\begin{array}{rclrclrcl}
E_1&=&\frac{i}{\sqrt{2}}(e_{13}-e_{24}),&\phantom{aaa}& e_{12}&=&\frac{-1}{\sqrt{2}}(E_6-iE_4)\\
E_2&=&\frac{1}{\sqrt{2}}(e_{13}+e_{24}),&\phantom{aaa}& e_{13}&=&\frac{1}{\sqrt{2}}(E_2-iE_1)\\
E_3&=&\frac{-i}{\sqrt{2}}(e_{14}+e_{23}),&\phantom{aaa}& e_{14}&=&\frac{-1}{\sqrt{2}}(E_5-iE_3)\\
E_4&=&\frac{-i}{\sqrt{2}}(e_{12}-e_{34}),&\phantom{aaa}& e_{23}&=&\frac{1}{\sqrt{2}}(E_5+iE_3)\\
E_5&=&\frac{-1}{\sqrt{2}}(e_{14}-e_{23}),&\phantom{aaa}&
e_{24}&=&\frac{1}{\sqrt{2}}(E_2+iE_1)\\
E_6&=&\frac{-1}{\sqrt{2}}(e_{12}+e_{34}),&\phantom{aaa}& e_{34}&=&\frac{-1}{\sqrt{2}}(E_6+iE_4)
\end{array}
$$
Then, the calculation gives the following result:
$$ f=-\sqrt{2} \Re(c)\left( \Im(\beta ) E_1+\Re (\beta) E_2-\Im (\alpha) E_3-\Im (c) E_4+\Re (\alpha) E_5+\Re (c) E_6\right) $$
Evidently there is a problem with this definition for $\Re (c)=0.$ But we are free to choose the scale factor in our definition, therefore we {\em define\,}:
\be f(c,U)\stackrel{df}{=} \Im(\beta ) E_1+\Re (\beta) E_2-\Im (\alpha) E_3-\Im (c) E_4+\Re (\alpha) E_5+\Re (c) E_6.\label{eq:iso}\ee
It is easy to see that the formula above provides an embedding of $U(1)\times SU(2)$ into the isotropic cone $\cN$ of $E^(4,2)$ that is transversal to the generator lines of $\cN,$ and therefore, by taking the quotient with respect to the multiplicative action of $\BR^+,$ a diffeomorphism from $U(1)\times SU(2)$ onto ${\hat P}\cN.$ Notice that we have $f(-c,-U)=-f(c,u),$ thus replacing $c\mapsto -c,$ $U\mapsto -U$ changes the orientation of the corresponding isotropic subspace.

Let us now return to the formula (\ref{eq:uu}) of sec. \ref{sec:ccms} that provides the embedding $\psi$ of $M$ into $U(2)$ via the Cayley transform. We rewrite it in the from
$ \psi(x)=cU',\, c\in U(1),\, U'\in SU(2),$
with
%calculated with inftya2
$ c=-(1+q(x)+2i x^0)/\sqrt{(1+q(x))^2+4{x^0}^2},$
$$ U'=\frac{1}{\sqrt{(1+q(x))^2+4{x^0}^2}}\bma 1-q(x)+2ix^3&2(ix^1+x^2)\\2(ix^1-x^2)&1-q(x)-2ix^3\ema.$$
Applying the formula (\ref{eq:iso}) we obtain
\begin{multline*}
 f(c,U')= \lambda\left(x^1E_1+x^2E_2+x^3E^3+x^0E_4\right)\\
 +\lambda\left(\frac{1}{2}(1-q(x))E_5-\frac{1}{2}(1+q(x))E_6\right),
\end{multline*}
where
$ \lambda= \frac{2\sqrt{2}}{(1+q(x))^2+4{x^0}^2}>0.$
This is the same map as the one given by Eq. (\ref{eq:tau}).
\subsubsection{From self--dual bivectors to the Clifford algebra and conformal spinors\label{sec:cliff}}
In Chapter 1.5.5.1 of \cite{angles} Pierre Angl\`{e}s generalizes earlier results of Deheuvels and shows how to embed the projective null cone of $E^{p,q}$ into the space of spinors of the Clifford algebra of this space. It is instructive to see how this method works in our case, yet in order to this we must first explicitly identify the space of spinors for our version of $E^{4,2}$ realized as self--dual bivectors in $H_{2,2}.$

\begin{lemma}
Define the following six complex matrices \be{\Gamma_\alpha}^i_k={\Sigma_\alpha}^{ij}Q_{jk},\, (\alpha=1,...,6;\,i,j,k=1,...,4)\label{eq:cli}\ee
and let $\Gamma_\alpha$ be the antilinear operators on $H_{2,2}$ defined by the formula:
$$({\Gamma_\alpha} f)^i = (\Gamma^\alpha)^i_j\,\overline{f^j},\quad f=(f^i)\in H_{2,2}.$$ Then the antilinear operators $\Gamma_\alpha$ satisfy the following anti--commutation relations of the Clifford algebra of $E^{4,2}:$
$$ \Gamma_\alpha\circ\Gamma_\beta +\Gamma_\beta\circ\Gamma_\alpha = 2\,Q_{\alpha_\beta}.$$
The space $H_{2,2}$ considered as an $8$--dimensional {\em real\,} vectors space carries this way an irreducible representation of the Clifford algebra $Cl_{4,2}.$ The Hermitian conjugation in $H_{2,2}$ coincides with the main anti--automorphism of $Cl_{4,2}.$ The space $H_{2,2}$ considered as a $4$-dimensional {\em complex\,} vector space carries a faithful irreducible representation of the even Clifford algebra $Cl^{+}_{4,2}.$\\
\label{lemma:bivectors}\end{lemma}
\begin{proof}
The formulas (\ref{eq:cli}) follow easily by a direct calculation. The first part of the Proposition follows then from the known fact that the Clifford algebra $Cl_{4,2}$ is isomorphic to the algebra $Mat(8,\BR),$  while the even Clifford $Cl^{+}_{4,2}$ is known to be isomorphic to $Mat(4,\BC)$ (cf. e.g., \cite[Table 1.1, p. 28]{angles}). Moreover, also by the direct calculation we have
$ (\Gamma_\alpha\circ\Gamma_\beta)^*=\Gamma_\beta\circ\Gamma_\alpha,$
which proves the statement about the main automorphism.
\end{proof}
\begin{proposition}The pseudo-Hermitian space $H_{2,2}$ is a spinor space for the Clifford algebra of its self--dual bivectors.\end{proposition}
\begin{proof}The proposition is an immediate consequence of Lemma \ref{lemma:bivectors}\end{proof}
In   \cite[Ch. 1.5.5.1, p. 44]{angles} Pierre Angl\'{e}s discusses a general method of embedding a projective quadric into the manifold of totally isotropic subspaces of a spinor space for the even Clifford algebra. Let us apply this method to our case adding at the same time a new element to this method. The original method can be described as follows: Consider $E^{p,q}$ as a vector subspace of its Clifford algebra $Cl_{p,q}.$ Let $S$ be a spinor space for $Cl^{+}_{p,q}$ endowed with its associated scalar product. For each non--zero isotropic vector $x\in E^{p,q}$ find another isotropic vector $y$ such that $2\langle x,y\rangle =1.$ Then $yx$ is an idempotent in $Cl^+_{p,q},$ and its kernel $S(x)$ is a totally isotropic subspace of $S$ that depends only on $x$ and not on $y.$ One disadvantage of this procedure in applications is that we are not being given a procedure for selecting $y$ for each given $x.$ This can be, however, in our case, easily improved.

Let us first describe the philosophy behind our procedure.\footnote{For more information cf.   \cite{born} and references therein.}} The set $D$ of maximal positive subspaces of $H_{2,2}$ is a complex symmetric domain for $U(2,2),$ $D=U(2,2)/(U(2)\times U(2)),$ and the manifold of maximal totally isotropic subspaces is its Shilov's boundary $\hat{D}.$ There is a one-to-one correspondence between maximal subspaces and Hermitian unitary operators $J$ in $H_{2,2}$ with the property that the scalar product $(x|Jy)$ is positive definite on $H_{2,2}.$ If $J$ is such an operator, then the associated maximal positive subspace is given by $\{z\in H_{2,2}:\,Jz=z\}.$ Every such $J$ is, in particular, an element of $SU(2,2),$ therefore it acts on its Shilov's boundary $\hat{D}.$ Acting on a given element of $\hat{D},$ it produces another element, its ``J-antipode''. We will take for $J$ the operator described by the matrix $G.$ It is then easy to see that in terms of isotropic vectors $E^{4,2}$ the corresponding action consists of flipping the signs of two coordinates: $(\bx,t,x^5,x^6)\mapsto (\bx,-t,x^5,-x^6).$ In other words - it corresponds to the action of the matrix $Q$ - cf. (\ref{eq:q}).

The geometrical idea described above, when implemented, leads to the following Proposition  \ref{prop:isot}.
\begin{proposition}
Let $x$ be a point in $M,$ $x=(x^0,\bx),$ let $X=\tau (x)$ be its image in $E^{4,2},$ as in Eq. (\ref{eq:tau}), and let $Y'=QX$ be its antipode. Let $Y=Y'/(2\langle X,Y'\rangle),$ so that $2\langle X,Y\rangle =1.$ Let $\hat{X}=X^1\Gamma_1+...+X^6\Gamma_6$ be the image of $X$ in $Cl_{4,2},$ and similarly for $\hat{Y}=Y^1\Gamma_1+...+Y^6\Gamma_6.$ Then $P=\hat{Y}\circ\hat{X}$ is an idempotent in the space $L(H_{2,2})$ of linear operators of $H_{2,2},$ whose kernel is a maximal totally isotropic subspace of $H_{2,2}$ consisting of vectors of the form
$\left(\begin{smallmatrix}Uv\cr v\end{smallmatrix}\right),$ where $U$ is the unitary matrix given by Eq. (\ref{eq:uu}).
\label{prop:isot}\end{proposition}
\begin{proof}
The proof follows by a straightforward though lengthy direct calculation.
\end{proof}
\section{Flat conformal structures\label{sec:fcs}}
 While the present paper concentrates on the Minkowski space, the results apply also to tangent space structures in more general case - they may also apply to conformally flat manifolds. In this section we will introduce the main concepts needed for such an extension and show that the embedding $\tau$ given by Eq. \ref{eq:tau} of section \ref{sec:ref2} can be understood geometrically by the {\it conformal development\,} with respect to the normal Cartan connection.
\subsection{The bundle $P^2(M)$}
 Let $M$ be a smooth $n$-dimensional manifold. Two maps  from open
neighborhoods of the origin $0\in\bR^n$ to $M$ define the same
$2$-jet at $0$ if and only if their partial derivatives up to the
second order coincide. The $2$-jet determined by such a map $e$ is
denoted $j^2_0(e).$ If $e$ is a diffeomorphism, then $j^2_0(e)$ is
called a second order frame at the point $p=e(0).$ The set of
all second-order frames is denoted by $P^2(M).$\footnote{For a somewhat different version cf. also \cite[pp. 138-152]{angles}}

Let $(x^\mu)$ be a local chart of $M$, and let $(t^a)$ be the
standard coordinates on $\bR^n.$ Given $j^2_0(e)$ such that $p$ is
in the domain of the chart, a set of coordinates of $j^2_0(e)$ is
defined by:
$$
\left\{\begin{array}{lcl} e^\mu&\doteq&x^\mu (p)\\
e^\mu_{\phantom{\mu}a}&\doteq&\frac{\partial(x\circ
e)^\mu}{\partial t^a }\vert_{t=0}\\
e^\mu_{\phantom{\mu}ab}&\doteq&\frac{\partial^2(x\circ
e)^\mu}{\partial t^a \partial t^b}\vert_{t=0}
\end{array}\right.
$$
If $(x^\mu)$ is replaced by $(x^{\mu\prime})$, the coordinates of
$j^2_0(e)$ change:
$$
\left\{\begin{array}{lcl} e^{\mu\prime}&=&x^{\mu\prime} (p)\\
e^{\mu\prime}_{\phantom{\mu}a}&=&\frac{\partial x^{\mu\prime}}{\partial x^\mu }(p)e^\mu_{\phantom{\mu}a}\\
e^{\mu\prime}_{\phantom{\mu}ab}&=&\frac{\partial
x^{\mu\prime}}{\partial x^\mu
}(p)e^\mu_{\phantom{\mu}ab}+\frac{\partial^2
x^{\mu\prime}}{\partial x^\mu
x^\nu}e^\mu_{\phantom{\nu}a}e^\mu_{\phantom{\mu}b}
\end{array}\right.
$$
It follows that $e^\mu_{\phantom{\mu}a}$ may be considered as an
ordinary (i.e., first order) frame at $p$. A natural projection
$P^2(M)\rightarrow P^1(M)$ exists, and is given by
$j^2_0(e)\mapsto j^1_0(e)$ or, in coordinates, by
$(e^\mu,e^\mu_{\phantom{\mu}a},e^\mu_{\phantom{\mu}ab})\mapsto
(e^\mu,e^\mu_{\phantom{\mu}a}).$ A simple interpretation can be
given to $\Eee{e}{\mu}{a}{b}.$ First notice that the matrix
$\Ee{e}{\mu}{a}$ is always invertible. Let $\Ee{e}{a}{\mu}$ denote
the inverse matrix, so that we have
$\Ee{e}{\mu}{a}\Ee{e}{a}{\nu}=\delta^\mu_\nu$ and
$\Ee{e}{a}{\mu}\Ee{e}{\mu}{b}=\delta^a_b .$ Define ``connection
coordinates of $e$'' by $$\Eee{e}{\mu}{\rho}{\sigma}\doteq
-\Ee{e}{r}{\rho}\Ee{e}{s}{\sigma}\Eee{e}{\mu}{\rho}{\sigma}.$$ It
follows from the transformation properties of the coordinates of
$e$ above that $\Eee{e}{\mu}{\rho}{\sigma}$ transform as
connection coefficients at $p.$ Therefore each section of $P^2(M)$
determines a pair: a section of $P^1(M)$ (i.e., a frame) and a
torsion-free affine connection on $M$, the correspondence being
bijective. In particular, if $P^1(M)$ is reduced to the orthogonal
or pseudo-orthogonal group, the Hilbert-Palatini principle for
General Relativity can be considered as a functional on the space
of sections of $P^2(M)$ Also notice that the diffeomorphisms group
of $M$ acts on $P^2(M)$ and on the space of its sections in a
natural way. If $e$ is a map from an open neighborhood of the
origin $0\in\bR^n$ to $M$, and if $\phi :M\rightarrow M$ is a
local diffeomorphism defined at $p=e(0),$ then $\phi\circ e$ is
another map from an open neighborhood of the origin $0\in\bR^n$
to $M.$ If $e_1$ and $e_2$ define the same second order frame:
$j^2_0(e_1)=j^2_0(e_2)$, then the composed maps define the same
second order frame as well: $j^2_0(\phi\circ e_1)=j^2_0(\phi\circ
e_2).$
\subsection{The structure group $G^2(n)$}
Let $G^2(n)$ denote the set of all second-order frames at
$0\in\bR^n.$ $G^2(n)$ is a group with the group multiplication law
given by
$j^2_0(h)j^2_0(h)\doteq j^2_0(h\circ k).$ The group $G^2(n)$
acts on $P^2(M)$ from the right
$j^2_0(e)j^2_0(h)\doteq j^2_0(e\circ h ).$
Corresponding to the canonical coordinates in $\bR^n$, there are
natural coordinates in $G^2(n)$:
$(\Ee{h}{a}{b},\Eee{h}{a}{b}{c}),$ and each $j^2_0(h)$ can be
uniquely represented  by the map $\bR^n\rightarrow\bR^n$ given by
$t^a\mapsto \Ee{h}{a}{r}t^r+\frac{1}{2} \Eee{h}{a}{r}{s}t^rt^s.$
In terms of natural coordinates the group composition law in
$G^2(n)$ can be written as
$(\Ee{h}{a}{b},\Eee{h}{a}{b}{c})(\Ee{k}{a}{b},\Eee{k}{a}{b}{c})=(\Ee{h}{a}{r}\Ee{k}{r}{b},
\Eee{h}{a}{r}{s}\Ee{k}{r}{b}\Ee{k}{s}{c}+\Ee{h}{a}{r}\Eee{k}{r}{b}{c})$
%%%%%%%%%%%%%
While the group $G^2(n)$ acts on $P^2(M)$ from the right, and
$P^2(M)$ is a principal bundle over $M$ with $G^2(n)$ as its
structure group, the group $Diff (M)$ of diffeomorphisms of $M$
acts on $P^2(M)$ from the left, by fibre preserving
transformations, commuting with the right action of $G^2(n)$ -
thus as an automorphism group of $P^2(M).$ An affine connection
can be considered as a section of a bundle associated to $P^2(M)$
via an appropriate representation of $G^2(n)$ by affine transformations.
\subsection{Reduction of $P^2(M)$ induced by a conformal structure}
Let now $M$ be an
orientable and oriented $n$--dimensional differentiable manifold.
Let $GL_+(n)$ be the group of $n\times n$ real matrices of positive
determinant. We denote by $TM$ the tangent bundle of $M,$ and by
$F_+$ the $GL_+(n)$ principal bundle of oriented linear frames of
$M.$ We denote by $\Lambda^n_+$ the bundle of oriented non-vanishing
$n$--vectors. $\Lambda^n_+$ is, in a natural way, a principal
$\BR_+$ bundle. Given a real number $w,$ let $V^w$ be the bundle
associated to $F^+$ via the representation $\rho_w$ of $GL^+(n)$ on
$\BR$ defined by $ \rho_w: GL^+(n)\ni A\mapsto \det(A)^{-w}\in
\BR.$  Since any
oriented frame $e$ defines an oriented $n$--vector
$e_1\wedge\ldots\wedge e_n,$ it follows that $V^w$ can be also
considered as the bundle associated to $\Lambda^n_+$ via the
representation $\BR_+\ni x\mapsto x^{w}\BR.$

Cross--sections of $V^w$ are called densities of weight $w.$ In what
follows we will use the ``hat'' symbol ${\hat {\phantom i}}$ to
distinguish densities from tensorial objects of weight $w=0.$ If
$e=\{e_i,\, i=1,\ldots,n\}$ is a frame at $p,$ and if $\hp$ is an
element in the fibre $V^r_p,$ then we denote by $\hp[e]$ the real
number representing $\hp$ with respect to the frame $e.$ We write
$\hp>0$ if $\hp[e]>0$ for some (and thus for every) oriented frame.
It follows from the very definition of the associated bundle that if
$A\in GL^+(n),$ then $\hp[eA]=\det[A]^{w} \hp[e].$

Let $r,s$ be a pair of real numbers, and let $\hp,{\hat \psi}$ be
positive densities of weight $w=r$ and $w=s$ respectively. Then
$(\hp{\hat \psi})[e]=\hp[e]{\hat \psi}[e]$ defines a density of
weight $w=r+s,$ while $\hp^s[e]=\hp[e]^s$ defines a density of
weight $w=rs.$

Let $x^\mu,\, \mu=1,\dots\,n$ be a local coordinate system on $M,$
and let $\partial_\mu$ be the basis made of vectors tangent to the
coordinate lines. Then a cross--section $\hp$ of $V^w$ is
represented by a real-valued function $\hp(x).$ When the local
coordinate system changes to another one, $x^{\mu'},$ then the
coordinate bases changes accordingly: $
\partial_{\mu'}=\frac{\partial x^\mu}{\partial
x^{\mu'}}\,\partial_\mu,$ and the corresponding numerical
representation of $\hp$ changes as follows: $
\hp'(x')=|\frac{\partial x'}{\partial x}|^{-w}\, \hp(x),$ where
$|\frac{\partial x'}{\partial x}|$ is the Jacobian of the coordinate
transformation.

By taking tensor products of tensor bundles with the line bundle
$V^w$ we can define, in an obvious way, tensor densities of weight
$w.$

Although much of what will follow
is true in a general case of an arbitrary (pseudo-)Riemannian
manifold, we will assume in the following that we are dealing
with the signature $(n-1,1),$ that our manifold $M$ is oriented and
time--oriented, and that all our local coordinate systems have
positive orientation and time--orientation.

Let $\eta_{ij}=\mbox{diag}\, (1,\ldots\,1,-1,\ldots ,-1),$ (signature $(p,q)$)and let
$0(\eta)$ be the subgroup of $GL(n)$ consisting of matrices
$\Lambda=(\Lambda^i{\phantom{i}j})\in GL(n)$ such that
$\Lambda^t\eta\Lambda=\eta,$ $\det \Lambda=1,$ and
let $SO_0(\eta)$ be the connected component of the identity in $O(\eta).$ By a (pseudo-) Riemannian structure on
$M$ we will mean a reduction of the $GL(n)$ principal bundle of the
linear frames of $M$ to $SO_0(\eta).$

There are several equivalent ways of defining a conformal structure
on $M.$ Probably the most intuitive way is to define it
as ``a Riemannian metric up to a scale''. Let $g$ and ${\tilde g}$ be
two metrics of $M.$ Then $g$ and ${\tilde g}$ are said to be
conformally related if there exists a positive function $\hp$ on $M$
such that $g(p)=\hp(p)g(p)$ for all $p\in M.$ Being ``conformally
related'' is, in fact, an equivalence relation, so that we can define
a conformal structure on $M$ as the equivalence class consisting of
conformally related metrics.

Let $C$ be a conformal structure on $M.$ For
any $g\in C,$ given a local coordinate system $x^\mu,$ we can define
$|g|$ to be the absolute value of the determinant $\det g_{\mu\nu},$
where $g_{\mu\nu}=g(\partial_\mu,\partial_\nu ).$ Then from the
transformation law: $ g_{\mu'\nu'}=\frac{\partial x^\mu}{\partial
x^{\mu'}}\frac{\partial x^\nu}{\partial x^{\nu'}}g_{\mu\nu}$ we
find that $ |g'|=|\frac{\partial x}{\partial x'}|^2\, |g|,$ so
that $|g|$ is a scalar density of weight $-2.$ Let us define $
\gamma_{\mu\nu}=\frac{g_{\mu\nu}}{|g|^{1/n}}.$ Then $\det
\gamma_{\mu\nu} = -1,$ $\gamma_{\mu\nu}$ is a symmetric tensor density of
weight $-2/n,$ and $\gamma_{\mu\nu}$ is independent of the choice of the
representative $g_{\mu\nu}$ in the conformal class $C$. In other
words: a conformal structure is uniquely characterized by a
symmetric tensor density of weight $-2/n,$ and signature $(p,q).$

Let $\tm M$ be the vector bundle of vector densities of weight
$w=1/n.$ Then, for any two vectors ${\hat u},{\hat v}\in \tm_p M$
the number $({\hat u},{\hat v})= \gamma_{\mu\nu}{\hat u}^\mu {\hat
v}^\nu$ is independent of the local coordinate system at $p$ - it
defines a bilinear form of signature $(p,q)$ on $\tm M.$ This
bilinear form characterizes uniquely the conformal structure $C.$

Let a conformal structure $C$ be given on $M.$ A general torsion--free affine connection which preserves $C$ is of the form
 $$\Gamma^\alpha_{\beta\gamma}={\hat \Gamma}^\alpha_{\beta\gamma}+\left(\delta^\alpha_\beta p_\gamma+\delta^\alpha_\gamma p_\beta-\gamma_{\beta\gamma}\gamma^{\alpha\rho}p_\rho\right),$$
where
 $\hat{\Gamma}^\alpha_{\beta\gamma}=\frac{1}{2}\left(\partial_\beta \gamma_{\gamma\rho}+\partial_\gamma \gamma_{\beta\rho}-\partial_\rho \gamma_{\beta\gamma}\right),$
and $\gamma^{\mu\nu}$ is the inverse matrix of $\gamma_{\mu\nu}.$ Therefore $P^2(M)$ can be reduced to $P^2_C(M)$ defined as consisting of second--order frames $e$ such that $(e^\mu_{\phantom{\mu}a}$ are conformal frames and $\Eee{e}{\mu}{\rho}{\sigma}$ are the coefficients of conformal connections. It is easy to see that the structure group $H$ of $P^2_C(M)$ is a subgroup of $G^2(n)$
consisting of pairs $(\Ee{h}{a}{b},\Eee{h}{a}{b}{c}),$ with $\Ee{h}{a}{b}\in CO_0(\eta),$ and
$ \Eee{h}{a}{b}{c}=\Ee{h}{a}{r}\left(\delta^r_b v_c+\delta^r_c v_b-\eta_{bc}\eta^{rs} v_s\right),$
where $CO_0(\eta)=SO_0(\eta)\times\BR^+,$ $v=(v_a)\in\BR^{n*}.$ It follows that $H$ is isomorphic to the semi--direct product $CO_0(\eta)\times\BR^{n*}$ with the multiplication law
$$ \left(\Ee{h}{a}{b},v_a\right)\left(\Ee{k}{a}{b},w_a\right)=\left(\Ee{h}{a}{r}\Ee{k}{r}{b},v_r\Ee{k}{r}{a}+w_a\right),$$
where $\Ee{h}{a}{b}=\exp (\sigma) \Ee{\Lambda}{a}{b},$ with $\exp (\sigma)\in\BR^+$ and $\Lambda\in CO_0(\eta).$ With $\left(\Ee{h}{a}{b},v_a\right)$ written as $(\theta,\Ee{\Lambda}{a}{b},v_a),$ one can easily verify that the following formula defines a representation $R$ of $H$ on $\BR^{n+2} =\BR^n\oplus\BR^2:$
$$ R(\theta,\Lambda,v)=\left(\begin{smallmatrix}\Ee{\Lambda}{r}{s}&\eta^{rd}v_s&\eta^{rs}v_s\\
\frac{-v_r}{\theta} &\frac{1+\theta^2-v^2}{2\theta}&-\frac{1-\theta^2+v^2}{2\theta}\\
\frac{v_r}{\theta}&-\frac{1-\theta^2-v^2}{1\theta}&\frac{1+\theta^2+v^2}{2\theta}
\end{smallmatrix}\right).$$
With $S=\left(\begin{smallmatrix}\eta&0&0\\0&1&0\\0&0&-1\end{smallmatrix}\right)$ we then have $R(\theta,\Lambda,v)^tSR(\theta,\Lambda,v)=S,$ therefore the representation $R$ realizes $H$ as a subgroup of the group $G=SO_0(p+1,q+1).$ The part of $G$ that is missing in $H$ is the translation group given by the following $SO_0(p+1,q+1)$ matrices $T(a),$ $a\in\BR^n:$
\be T(a)=\left(\begin{smallmatrix}\delta^r_s&-a^r&a^r\\\eta_{rs}a^s&1-a^2/2&a^2/2\\ \eta_{rs}a^s&-a^2/2&1+a^2/2\end{smallmatrix}\right),\label{eq:ta}\ee
- Cf. section \ref{sec:translations}.
 The Lie algebra generators $so(p+1,q+a)$ take now the following form:
$$ D=\frac{d D(\exp(\sigma),E,0)}{d\sigma}|_{\sigma=0}=\left(\begin{smallmatrix}0&0&0\\0&0&1\\0&1&0\end{smallmatrix}\right),\quad
\frac{1}{2}\Ee{\omega}{r}{s}\Ee{M}{s}{r}=\left(\begin{smallmatrix}0&0&0\\0&\Ee{\omega}{r}{s}&0\\0&0&0\end{smallmatrix}\right)$$
$$ v_rK^r=\left(\begin{smallmatrix}0&\eta^{rs}v_s&\eta^{rs}v_s\\-v_r&0&0\\v_r&0&0\end{smallmatrix}\right),\quad
w^rP_r=\left(\begin{smallmatrix}0&-w^r&w^r\\\eta_{rs}w^s&0&0\\\eta_{rs}w^s&0&0\end{smallmatrix}\right).$$
\subsection{The enlarged conformal bundle and the normal Cartan connection}
With $H$ being a subgroup of $G,$ as above, we can build now the associated bundle $\tilde{P}^2_C(M)=P^2_C(M)\times_H G,$ which is a principal $G$-bundle (cf. e.g., \cite[p. 4]{leitner} and references therein). If $n=p+q\geq 3,$ then this new bundle is naturally equipped with a principal connection, the normal Cartan connection, which can be described as follows.\\
Let $g$ be a metric in the conformal class $C,$ let $e_a$ be an (local) orthonormal frame of $g,$ and $R$ its curvature tensor. Then, in a coordinate system $x^\mu,$ the covariant derivative $\nabla_\mu Z$ of a section $Z$ of the associated vector bundle $\tilde{P}\times_R E^{p+1,q+1}$ is given by the following expression - cf. e.g., \cite[Ch. 4.4]{kobayashi},\cite[p. 14]{leitner},\cite[p. 196]{angles} :
$$ \nabla_\mu Z=\partial_\mu Z+\Gamma_\mu Z,$$
with
$ \Gamma_\mu =\frac{1}{2}\Gamma_\mu^{rs}\Ee{M}{s}{r}+\frac{1}{n-2}\left(R_{\mu\sigma}-\frac{1}{2(n-1)}Rg_{\mu\sigma}\right)K^\sigma-P_\mu,$
where $K^\mu=e_r^\mu,$ and $P_\mu=e^r_\mu P_r.$

In a natural way we can then build the associate bundle $\tilde{P}\times_G E^{p+1,q+1}$ with $E^{p+1,q+1}$ as a typical fibre, and we can constructed the projective quadric $\tilde{M}_x$ at each point $x\in M.$

Now, suppose $M$ is connected and simply connected and the conformal structure is flat. In this case we can choose (cf. \cite[Ch. I.2]{kobayashi}) $g_{\mu\nu}=\eta_{\mu\nu}.$ The covariant derivative $\nabla_\mu Z$ reduces in this case to
$ \nabla_\mu Z=\partial_\mu Z-P_\mu Z.$ In an adapted coordinate system $x^\mu$ we choose the ``origin'' of the ``compactified tangent space'' to correspond to the point $(0,\frac12,-\frac12)$ of $E^{p+1,q+1}.$ Connecting the point $x\in M$ with $0\in M$ by the path
$x(t)=(1-t)x$ we can then transport parallely the origin $(0,\frac12,-\frac12)$ at to the point $0\in M.$ The parallel transport rule gives
us
$ 0=DZ(x(t))/dt= dZ(x(t))/dt-dx^\mu /dt P_\mu Z(x(x(t)),$
or, in our case,
$dZ/dt=-x^\mu P_\mu Z,$
which solves to
$ Z(1)=\exp(x^\mu P_\mu)Z(0),$
or, applying Eq. (\ref{eq:ta}):
$ Z(1)=(x,(1-x^2)/2),-(1+x^2)/2),$
which is nothing but the standard embedding (\ref{eq:tau}).
\section{Concluding remarks}
This paper has provided a mathematical analysis of algebraic and geometrical aspects of the Minkowski space compactification. Some omissions, faulty reasoning and lack of precision in the existing literature dealing with this subject has been pointed out and analyzed in some detail. In addition to the standard compactification by adding a ``light cone and a 2-sphere at infinity'' also its double covering isomorphic to $U(1)\times SU(2)$ has been discussed. A pictorial representation has been proposed and the corresponding ``Penrose diagrams'' have been derived. The role of the conformal inversion and the representation of null geodesics has been touched upon as well. Applications to flat conformal structures, including the normal Cartan connection and conformal development has been discussed in some detail. In  appendix \ref{app:app1} a detailed discussion of the spaces of null lines in a general case  of a pseudo--Hermitian space $H_{p,q}$ has been given.
\section{Acknowledgments}
Thanks are due to Pierre Angl\`{e}s for his encouragement, for critical reading of the manuscript and for many constructive discussions. I also thank  Rafa\l \ Ab\l{}amowicz for the discussion and many suggestions concerning the form and the content of this paper. Thanks are due to Don Marolf for pointing out a possible usefulness of adding the analysis of the $IO(3,1)$ action, also to Alexander Levichev for useful suggestions. Special thanks are also due to Pawe{\l} Nurowski for sharing with me some of his knowledge and for pointing to me the question discussed in appendix \ref{app:app1}. I am indebted to Nikolay M. Nikolov for sending me some of his papers and for discussion.
\appendix
\appendixpage
\section{Killing vector fields for the left action of $U(2)$ on itself\label{app:app1}}
\subsection{The problem}
We take the group $U(2)$ in the standard matrix form. It has the manifold structure of $(S^1\times S^3)/Z_2$ - the same as the compactified Minkowski space.

Now, let $\omega$ be the Maurer-Cartan form of $U(2),$  a $2\times 2$  matrix of one-forms. Taking the determinant of $\omega$ with understanding that the multiplication of one-forms is to be understood as a symmetrized tensor product, we obtain a symmetric bilinear form $ g=\det (\omega).$ This form is non-degenerate of Lorentzian signature and is conformal to the flat Minkowski metric under the standard
identification of $U(2)$ as the compactification of the Minkowski space $M$. The metric $g$ obtained this way is, by its very construction, invariant under the left action of $U(2)$ on itself. Therefore the left action of $U(2)$ on itself leads to conformal transformations of $M.$

Precisely which subgroup of the conformal group corresponds to this left action of $U(2)$ on itself?
\subsection{The solution}
It is well known that the group $SU(2,2)$ acts by conformal automorphisms on the compactified Minkowski space (see e.g., \cite{wk}). The group $U(2,2)$ consists of block matrices
$\left(\begin{smallmatrix}A&B\cr C&D\end{smallmatrix}\right)$ with entries $A,B,C,D$ which are $2\times 2$ complex matrices satisfying the relations $A^*A-C^*C=D^*D-B^*B=E$ and $A^*B-C^*D=0.$
Its action on $U(2)$ is given by the fractional linear transformations:
\be U\mapsto U'=(AU+B)(CU+D)^{-1},\label{eq:action}\ee
with $CZ+D$ being automatically invertible for $U\in U(2).$
By specifying $B=C=0,\, D=E,$ we see that $A$ is in $U(2).$ Therefore the left action of $U(2)$on itself is a particular case of the linear fractional transformations as above.

In order to describe these transformations in the Minkowski space, we can use the Cayley transform as in   \cite{uhlmann63}. Or, we can inverse Cayley-transform the matrices of $U(2,2)$ and act on the Minkowski space represented by hermitian $2\times 2$ matrices in the standard form:
$ X=x^\mu\sigma_\mu,$ where
$ \sigma_0=\left(\begin{smallmatrix}1&0\cr 0&1\end{smallmatrix}\right),\;\sigma_1=\left(\begin{smallmatrix}0&1\cr 1&0\end{smallmatrix}\right),\;\sigma_2=\left(\begin{smallmatrix}0&-i\cr i&0\end{smallmatrix}\right),\; \sigma_3=\left(\begin{smallmatrix}1&0\cr 0&-1\end{smallmatrix}\right)$
The action of $U(2,2)$ is still described by fractional linear transformations
\be X\mapsto X'=(RX+S)(TX+Q)^{-1},\label{eq:action2}\ee
where (cf. \cite[(2.16)]{ruhl})$A=\frac12(R+iS-iT+Q),$
$B=\frac12(-R+iS+iT+Q),$
$C=\frac12(-R-iS-iT+Q),$
$D=\frac12(R-iS+iT+Q).$ With $B=C=0$ and $D=E$ we easily find that
$R=\frac12(A+E),$
$Q=\frac12(A+E),$
$S=-i\frac12(A-E),$
$T=i\frac12(A-E).$ Consider now a one-parameter subgroup $A(\tau)$ of $U(2).$ By differentiating the equation
$ X(\tau)=(R(\tau)X+S(\tau))(T(\tau)X+Q(\tau))^{-1}$
at $\tau=0,$ and putting $A(0)=E,$ ${\dot A}(0)=i\sigma$ we obtain:
$ {\dot X}=\frac{i}{2}(\sigma X-X\sigma)+\frac{1}{2}\sigma+\frac{1}{2}X\sigma X.$
Denoting by $Z_\mu$ the vector fields corresponding to $\sigma=\sigma_\mu$ we easily find their components using the simple algebra:
$ (Z_\mu)^\nu= \frac{1}{2}\mbox{tr} ({\dot X}\sigma_\nu ).$
The result is as follows:
$$Z_0=\frac{1}{2} \left(1+t^2+x^2+y^2+z^2\right)\partial_t+t(x\partial_x+y\partial_y+z\partial_z),$$
$$Z_1=tx\,\partial_t+\frac{1}{2} \left(1+t^2+x^2-y^2-z^2\right)\partial_x+(xy+z)\partial_y+(xz-y)\partial_z$$
$$Z_2=ty\,\partial_t+(xy-z)\partial_x+\frac{1}{2} \left(1+t^2-x^2+y^2-z^2\right)\partial_y+(xz+y)\partial_z$$
$$Z_3=tz\,\partial_t+(xz+y)\partial_x+(yz-x)\partial_y+\frac{1}{2} \left(1+t^2-x^2-y^2+z^2\right)\partial_z.$$
We can now compare these vector fields with the formulas for the standard generators $P_\mu,\,K_\mu,\, M_{\mu\nu}$ of the conformal group as given, for instance, in   \cite{mn}:
$$
P_\mu=-\partial_\mu,\,
M_{\mu\nu}=x_\mu\partial_\nu-x_\nu\partial_\mu,\,
K_\mu=-2x_\mu(x^\nu\partial_\nu )+x^2\partial_\mu.$$
By an easy calculation we find:
$Z_0=\frac{1}{2}(K_0+P_0),\quad
Z_i=\frac{1}{2}(K_i-P_i)+L_i,\quad (i=1,2,3),$
where $L_i=\epsilon_{ijk}M_{jk}.$

\newpage
\begin{figure}[!ht]
\begin{center}
      \includegraphics[width=5cm, keepaspectratio=true]{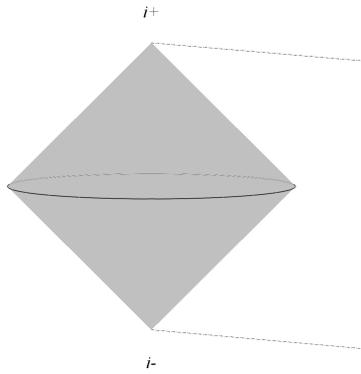}
\end{center}
  \caption{Pictorial representation of the conformal infinity with one dimension skipped. Double light cone at infinity with endpoints identified. While topologically correct this representation is misleading as it suggests non differentiability at the base, where the two half-cones meet.}
\label{fig:fig3}\end{figure}
\begin{figure}[!hb]
\begin{center}
     \includegraphics[width=5cm, keepaspectratio=true]{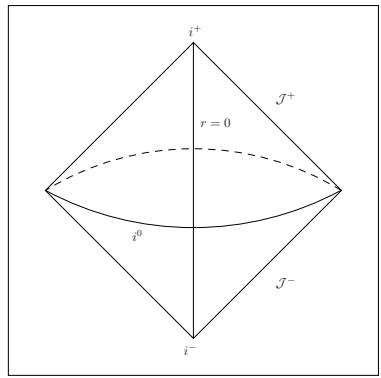}
\end{center}
  \caption{Conformal infinity of Minkowski space from   \cite[p. 178]{beem}. The meaning of this picture is quite different from the one in Fig. \ref{fig:fig3}, where the points $i^0,i^+,i^-$ are identified. The 2-sphere indicated in the middle of this picture is just one point $i^0$ and {\bf not} the true 2-sphere of Fig. \ref{fig:fig3}}
\label{fig:fig3a}\end{figure}
\begin{figure}[ht!]
\begin{center}
%    \leavevmode
      \includegraphics[width=5cm, keepaspectratio=true]{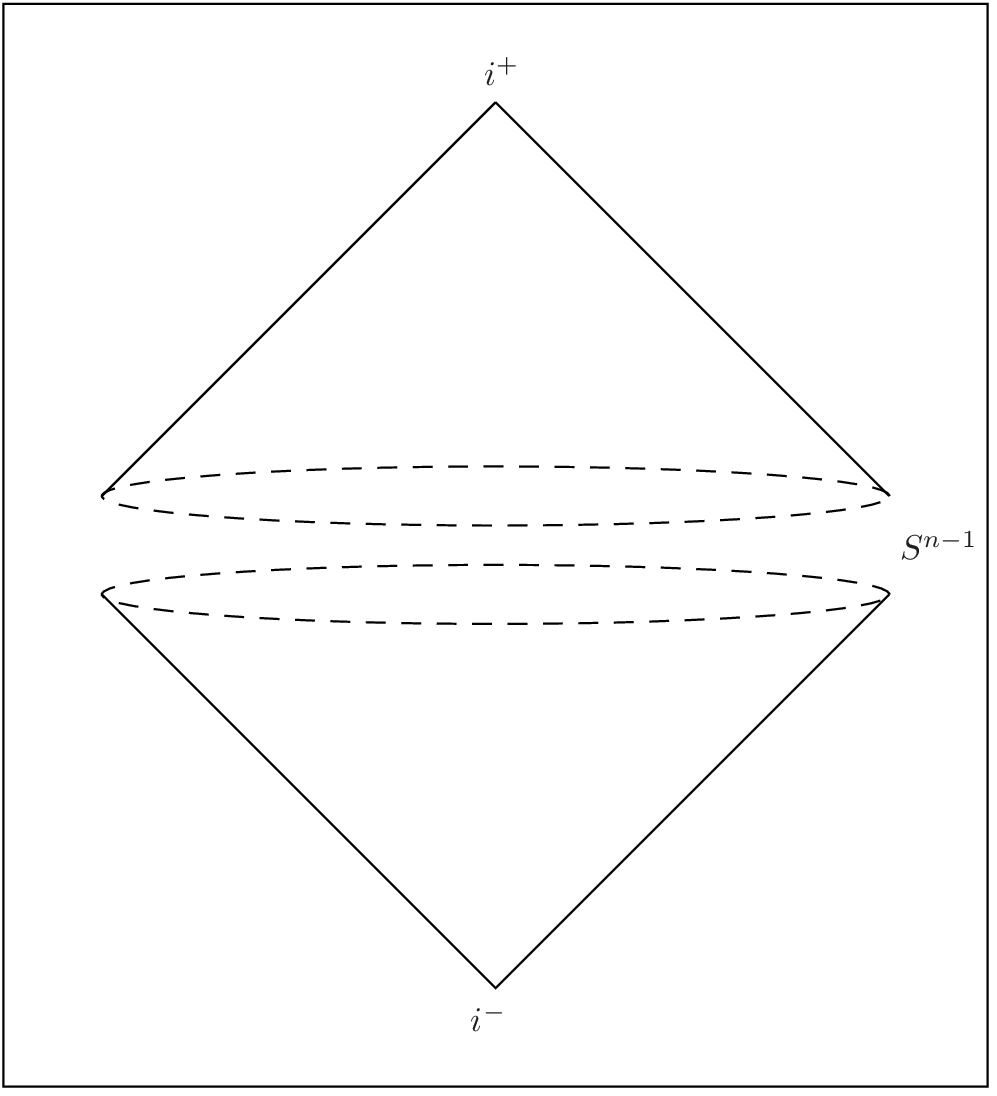}
\end{center}
  \caption{Chronological boundary for $\hbox{\ddpp L}^{n+1}$ - Figure. 2 from   \cite{flores}}
\label{fig:fig3b}\end{figure}
\begin{figure}[hb!]
\begin{center}
    \leavevmode
      \includegraphics[width=5cm, keepaspectratio=true]{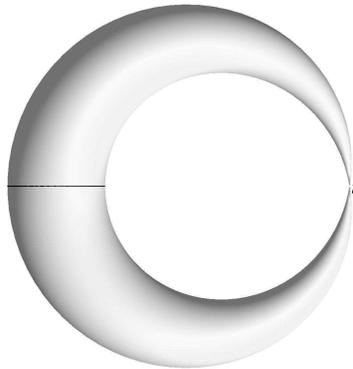}
\end{center}
  \caption{Differentiably correct pictorial representation of the conformal infinity with one dimension skipped. A torus squeezed to a point $I^+=I^-=I^0$ at $\psi=0.$ All null geodesics described in this section pass through this point.}
\label{fig:fig4}\end{figure}
\begin{figure}[h!]
\begin{center}
    \leavevmode
      \includegraphics[width=5cm, keepaspectratio=true]{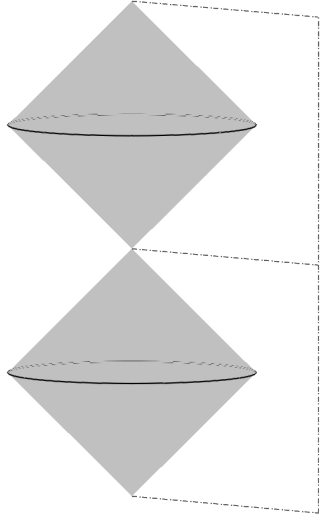}
\end{center}
  \caption{Pictorial representation of the double covering conformal infinity. Double double light cone. Points connected by a dashed line are, in fact, a one point. This representation is also topologically correct but differentiably misleading.}
\label{fig:fig5}\end{figure}
\begin{figure}[h!]
\begin{center}
    \leavevmode
      \includegraphics[width=5cm, keepaspectratio=true]{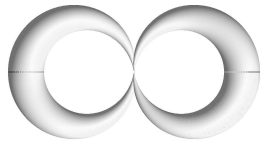}
\end{center}
  \caption{Pictorial representation of the double covering conformal infinity. A pair of tori squeezed at a common point.}
\label{fig:fig6}\end{figure}
\end{document}